\documentclass[fleqn,10pt]{cas-dc} 

\usepackage{wasysym} 
\usepackage{amsfonts,amssymb}
\usepackage{amsmath}
\usepackage{rotating}
\usepackage{pdflscape,array,booktabs}
\usepackage{tablefootnote}
\usepackage{threeparttable}
\newcolumntype{P}[1]{>{\raggedright}p{#1}}
\newlength{\ninetabcol}
\newlength{\eighttabcol}
\newlength{\sixtabcol}
\newlength{\fivetabcol}
\usepackage[mathlines]{lineno}

\usepackage{url} 
\usepackage{caption} 
\usepackage{subcaption} 
\usepackage[authoryear]{natbib}

\def\tsc#1{\csdef{#1}{\textsc{\lowercase{#1}}\xspace}}
\tsc{WGM}
\tsc{QE}

\begin{document}
\let\WriteBookmarks\relax
\def\floatpagepagefraction{1}
\def\textpagefraction{.001}
\widowpenalty=0
\clubpenalty=0

\shorttitle{Magnitude of stable iron isotope fractionation limited by multiple stages of terrestrial core formation}    

\shortauthors{Nathan et al.}  

\title [mode = title]{Magnitude of stable iron isotope fractionation limited by multiple stages of terrestrial core formation}

\author[1]{Gabriel Nathan}[type=author,orcid=0000-0002-4261-6677]

\cormark[1]


\ead{nathanga@msu.edu}

\address[1]{Dept. of Earth \& Environmental Sciences, Michigan State University, East Lansing, MI 48824, USA}

\author[1]{Seth A. Jacobson}[type=author,orcid=0000-0002-4952-9007]            

\begin{abstract}
One proposed mechanism for generating iron isotopic differences between planetary mantles and chondrites is metal-silicate equilibration during terrestrial core formation.
Prior studies of this isotopic fractionation effect employ single-stage core formation models that are inconsistent with reproducing the siderophile element budget of the Earth's mantle.
Here, we model iron isotopic evolution of the Earth's mantle in a multistage core formation scenario that is consistent with dynamic models of planet formation and reproduces the geochemistry of the bulk silicate Earth, specifically the refractory moderately siderophile elements.
We find that multiple stages of core formation rebalance the fractionating effect of metal-silicate equilibration for the iron isotopic system, comparable to or smaller than typical analytical uncertainties in mantle rocks.
Ultimately, this rebalancing effect - coupled with stochastic accretion and differentiation histories which further complicate planetary formation processes - makes it unlikely that isotopic fractionation during metal-silicate equilibration is the primary mechanism responsible for the iron isotopic composition of the Earth's mantle.
\end{abstract}

\begin{keywords}
Planetary differentiation \sep Core formation \sep iron isotopic composition 
\end{keywords}

\maketitle

\section{Introduction}
\label{introduction}
The abundance of iron, its speciation, and isotopic ratios are crucial tools to decipher the origin and evolution of a wide range of solar system materials.
Iron isotopes of chondritic meteorites show no statistical difference from the reference standard IRMM-014 \citep{sossi2016ironsystematics}, but there is deviation from the reference composition in Earth materials \citep[e.g.,][]{poitrasson2004iron} and other meteoritic groups such as angrites \citep[e.g.,][]{wang2012iron}.
While the root cause of heterogeneity among these planetary bodies' iron isotopic compositions is unclear, several mechanisms have been proposed, including vapor loss \citep[e.g.,][]{sossi2016ironmars}, partial melting \citep{weyer2007iron}, disproportionation of iron in the mantle \citep{williams2012oxidation}, and core formation \citep[e.g.,][]{polyakov2009iron,elardo2017iron,ni2022planet}.
Core formation is an appealing mechanism to generate isotopic heterogeneity in iron because all planetary bodies of large enough size experience this process and iron is the primary constituent of metallic cores in terrestrial planetary bodies \citep[e.g.,][]{elardo2017iron,ni2022planet}.
Extensive measurements of iron isotopic fractionation due to the metal-silicate equilibration show complex behavior of iron isotopes that partition between phases differently in response to pressure, temperature, and the presence of other elements \citep[e.g.,][]{poitrasson2009iron, hin2012iron, shahar2015sulfur}.

Here, we model iron isotopic fractionation due to metal-silicate equilibration in the context of multistage accretion and differentiation of terrestrial planets in order to assess the hypothesis that planetary core formation is a significant source of iron isotopic fractionation.
While other mechanisms of iron isotopic fractionation have been proposed, we focus this study on core formation in particular. 
Prior studies of iron isotopic fractionation have relied on simplified single-stage core formation models \citep[e.g.,][]{elardo2017iron,ni2022planet}.
Single-stage models assert that the entire mantle and core of the Earth fully equilibrated with one another, and did so at mid-mantle pressures and temperatures (as opposed to the pressure-temperature conditions of the core-mantle boundary). 
This problematic assumption produces a differentiation scenario which is incompatible with both astrophysical models of planetary accretion (which indicate planetary impacts occur frequently during accretion) as well as with geochemical constraints, which show difficulty in reproducing the abundances of moderately siderophile elements in the bulk silicate Earth (BSE) at a single pressure-temperature condition.
Some models otherwise assert that the oxygen fugacity of the Earth followed a specific trend that may not be compatible with the evolving chemistry of the Earth.
By modeling multi-stage core formation and using chemical mass-balance to calculate evolving fugacity conditions, we are able to self-consistently simulate Earth's iron isotopic fractionation during its accretion.
This approach allows comprehensive characterization the behavior of the iron isotopic system during Earth's formation and assess the likelihood of core formation as a source of iron isotopic heterogeneity in the terrestrial planets.

\begin{figure*}[!t]
    \centering
    \includegraphics[width=0.8\textwidth]{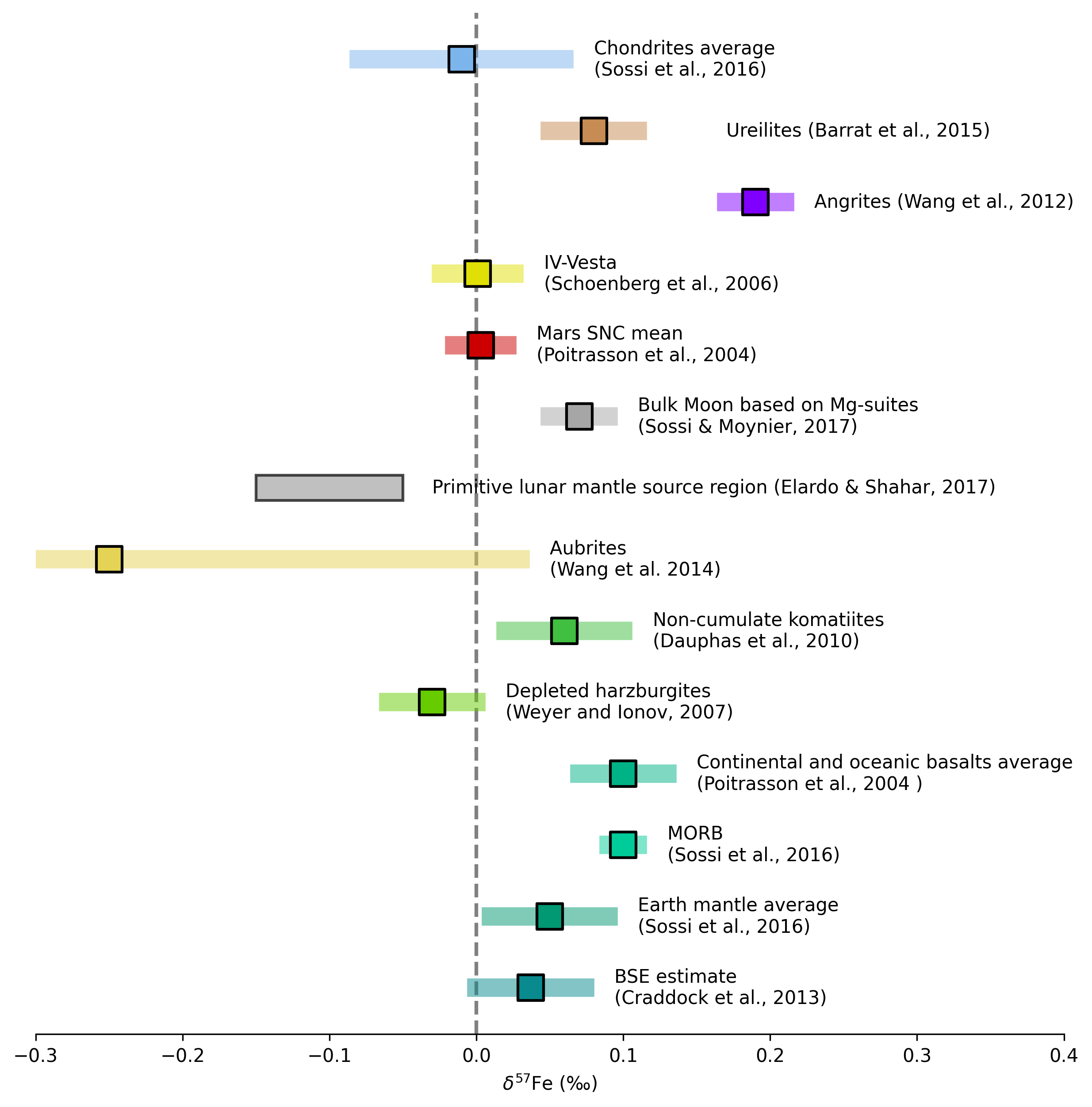}
    \caption{
    Literature $\delta^{57}\text{Fe}$ isotopic compositions of notable solar system materials are shown on the horizontal axis with the mean estimate and standard deviation, ordering of materials on the vertical axis is for visual aid.
    In particular, we note that terrestrial, lunar, martian and Vestan materials are (mostly) isotopically heavier than or consistent with a chondritic iron isotope composition.
    Primitive lunar mantle source region values are reported as a range from \citet{elardo2017iron}.
    Green colors represent terrestrial materials (average and specific assemblages).
    Shaded regions around markers display either error (2 S.E.) of measurements or standard deviation (1 SD) of compilations of values originally reported in cited literature.
    Compositions from \protect\citet{sossi2016ironsystematics, barrat2015iron, wang2012iron, schoenberg2006modes, sossi2017iron, elardo2017iron, wang2014iron, dauphas2010komatiites, weyer2007iron, poitrasson2004iron, craddock2013abyssal}}.
    \label{fig:fe_compositions}
    
\end{figure*}

Measurements of iron isotopic composition in various solar system samples display heterogeneity between and within planetary bodies, presenting a complicated state of solar system iron isotopic variations, as summarized in Figure \ref{fig:fe_compositions}.
Iron isotopic differences between solar system samples are displayed in delta notation, where $\delta^{i}\text{Fe}$ represents how much the Fe isotopes of 56, 57, or 58 in a sample deviate from a standard reference value in parts per thousand: $\delta^{i}\text{Fe}_{sample} = \left([^{i/54}\text{Fe}]_{sample}/[^{i/54}\text{Fe}]_{standard} - 1\right) \times 1000$.
Carbonaceous, ordinary, and enstatite chondrite meteorites have $\delta^{57}\text{Fe} = -0.01 \pm 0.07\permil$ (1SD, from the \citet{sossi2016ironsystematics} literature survey of \citet{craddock2011iron, needham2009ordinary, wang2014iron, schoenberg2006modes}) and the chondritic composition of $^{56}\text{Fe}/^{54}\text{Fe}$ is nearly identical to the reference standard IRMM-014 \citep{craddock2011iron}.
Martian meteorites have a mean composition $\delta^{57}\text{Fe} = 0.003 \pm 0.018\permil$ \citep{poitrasson2004iron} and an extrapolation based on Mg\# suggests Mars' mantle has $\delta^{57}\text{Fe} = -0.04 \pm 0.03\permil\text{ (1SD)}$ \citep{sossi2016ironmars}.
Vesta is also similarly close to a chondritic composition; eucrite meteorites suggest the silicate component of the HED (howardite-eucrite-diogenite) parent body had an average $\delta^{56}\text{Fe} = -0.001 \pm 0.017\permil$ \citep{schoenberg2006modes}.
Angrites are enriched in heavy isotopes compared to chondrites, with $\delta^{57}\text{Fe} = +0.192 \pm 0.014\permil \text{ (95\% confidence interval)}$ \citep{wang2012iron}.
Ureilites are also isotopically heavier than chondrites with $\delta^{56}\text{Fe} = +0.056 \pm 0.008\permil \text{ (2SD)}$ \citep{barrat2015iron}.
The aubrites (enstatite achondrites) are isotopically lighter ($\delta^{56}\text{Fe}=-0.170 \pm 0.189\permil \text{ (2SD)}$ \citep{wang2014iron}) but are not a homogeneous population and are not directly comparable to other homogeneous planetary reservoirs.
Lunar material has the same isotopic composition as terrestrial material, within uncertainty both for $\delta^{57}\text{Fe} \sim 0.1 \permil$ \citep{poitrasson2019reassessment, sossi2017iron} as well as for nucleosynthetic anomaly $\mu^{54}\text{Fe}$ \citep{hopp2025inner}.

Within Earth materials, there is variability in iron isotopic composition, and so there is difficulty in determining a representative iron isotopic composition of the bulk silicate Earth (BSE), as shown in Fig.~\ref{fig:fe_compositions}.
Compared to chondritic composition, continental and oceanic basalts are nearly uniformly isotopically heavier in iron $\delta^{57}\text{Fe} = +0.10\pm0.03\permil$ \citep{poitrasson2004iron, schoenberg2006modes,weyer2005iron}, including mid-ocean ridge basalts (MORB), which show a signal of $\delta^{57}\text{Fe} = +0.10 \pm 0.01\permil$ \citep{sossi2016ironsystematics}.
This elevated composition of basaltic rock has been interpreted to indicate that the bulk silicate Earth (BSE) may be slightly heavier than chondritic \citep[e.g.,][]{poitrasson2004iron}, despite metasomatism causing a large amount of Fe isotopic heterogeneity $>0.9\permil$ and the large spread of mantle xenolith composition \citep{poitrasson2013iron}.
For instance, \citet{poitrasson2004iron} argues the bulk silicate Earth has $\delta^{57}\text{Fe} \approx +0.1\permil$ relative to IRMM-14. 
\citet{sossi2016ironsystematics} argues that the composition of lherzolites (including continental xenoliths, abyssal peridotites, and orogenic massifs, but not basalts) suggests a heavier-than-chondritic iron composition for the BSE $\delta^{57}\text{Fe} = +0.05\pm 0.01\permil$.
The origin of this putative heavy iron composition in Earth's mantle is not known.
In contrast, \citet{craddock2013abyssal} contends that the Earth mantle average is chondritic within errors ($\delta^{56}\text{Fe} = +0.025 \pm 0.025\permil$, which implies $\delta^{57}\text{Fe} = +0.037 \pm 0.037\permil$, assuming mass-dependent fractionation behavior).

The processes that set the major and trace element compositions of the Earth (e.g., core formation, partial melting, and metasomatism) likely cause isotopic fractionation as well and so are natural candidates for Earth's heavy iron.
Of these, core formation is frequently invoked as a primary driver of planetary-scale isotopic evolution \citep[e.g.,][]{polyakov1994fractionation,Georg2007,nielsen2020vanadium} and isotopic fractionation of the iron system in particular \citep{roskosz2006experimental,polyakov2009iron,shahar2015sulfur,shahar2016iron,elardo2017iron,elardo2019iron,ni2022planet}.
Metal-silicate equilibration during terrestrial core formation is a particularly appealing mechanism to explain solar system iron isotopic variation for two primary reasons.
First, core formation is universal: every terrestrial body that experiences large scale melting undergoes metal-silicate differentiation, resulting in the segregation of a central iron core surrounded by a silicate rich mantle.
Second, iron is the dominant element within a planet's core yet considerable iron remains in the silicate mantle, i.e., iron behaves as a moderately siderophile element.

The equilibration of metal and silicate material during core formation may result in iron isotopic fractionation because of bond strength differences between immiscible liquid metal and liquid silicate phases \citep{urey1947thermodynamic,bigeleisen1947calculation,young2002kinetic}.
Iron isotopic fractionation after metal-silicate equilibration has been studied extensively in the laboratory via direct measurement of quenched metal and silicate after equilibration at high pressures and temperatures \citep[e.g.,][]{elardo2017iron,kubik2022absence}. 
It is also determined indirectly via nuclear resonant inelastic X-ray spectroscopy (NRIXS) which extracts iron force constants by processing nuclear resonance spectra \citep[e.g.,][]{polyakov2009iron, liu2017iron, ni2022planet}.
Alternatively, \textit{ab initio} estimates of iron isotopic fractionation during metal-silicate equilibration can be made computationally \citep[e.g.,][]{luo2024iron}.
These different measurement methods produce varying conclusions; some direct experimental results show a metallic phase that is isotopically light following metal-silicate equilibration \citep[e.g.,][]{roskosz2006experimental}, while others show no iron fractionation during metal-silicate equilibration \citep{hin2012iron}.
The magnitude of measured fractionation in these experimental results can change with the inclusion of light elements such as sulfur \citep[e.g.,][]{shahar2015sulfur} or variation in pressure \citep{shahar2016iron}. 
Some NRIXS results suggest that core formation is responsible for Earth's heavy Fe isotopic composition \citep{polyakov1994fractionation}.
See \citet{shahar2020assessment} for a thorough review of experimental methods and results, which are only summarized below.

\begin{figure*}[!htbp]
    \centering
    \includegraphics[width=\textwidth]{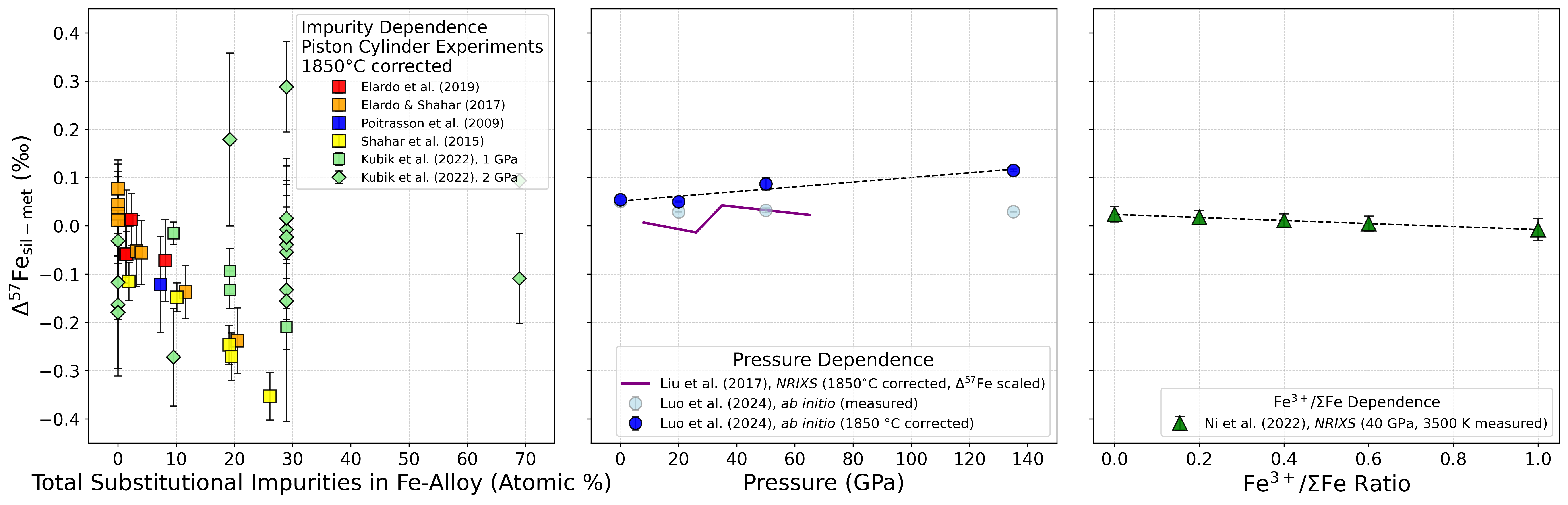}
    \caption{
        $\Delta^{57}\text{Fe}_\text{sil-met}$ variation depends on substitutional impurities, pressure (GPa), and $\text{Fe}^{3+}/ \sum\text{Fe}$.
        Impurity-dependence data from a number of piston–cylinder experiments are shown in the left hand subfigure \citep{elardo2019iron, elardo2017iron, poitrasson2009iron, shahar2015sulfur,kubik2022absence}. $\Delta^{57}\text{Fe}_\text{sil-met}$ are corrected to 1 GPa and 1850~°C, except \citet{kubik2022absence} measurements which are corrected to 1850~°C but measured at 1 GPa or 2 GPa, as indicated.
        Pressure-dependence values are displayed in the center subfigure, with data compiled from \textit{ab initio} \citep{luo2024iron} and NRIXS \citep{liu2017iron} studies.
        Plotted values are corrected to temperatures of 1850~°C and from $\Delta^{56}\text{Fe}_\text{sil-met}$ to $\Delta^{57}\text{Fe}_{sil-met}$, for ease of comparison and consistency.
        Measured values from \citet{luo2024iron} without corrections are shown in transparent blue markers.
        $\text{Fe}^{3+}/ \sum\text{Fe}$ dependence values are displayed in the right hand subfigure with data from NRIXS measurements \citep{ni2022planet}.
        Values from \citet{ni2022planet} at conditions of 40 GPa and 3500 K are shown in green markers.
        Trend lines shown in the middle and right hand panel are shown for visual aid.
        Error bars for NRIXS and \textit{ab initio} measurements are shown but are typically smaller than displayed size of marker.
        These figures are similar to those in Figure 7 in \citet{shahar2020assessment}.
    }
    \label{fig:capDelta57Fe_three_panel}
\end{figure*}

Laboratory efforts to quantify the effects of metal-silicate equilibration have evolved due to advancing techniques and ongoing debate about the relevant compositions for metal and silicate phases.
With a couple of notable exceptions \citep{poitrasson2009iron,hin2012iron,kubik2022absence}, the weight of this laboratory evidence \citep{roskosz2006experimental,polyakov2009iron,shahar2015sulfur,shahar2016iron,elardo2017iron,liu2017iron,elardo2019iron,ni2022planet,luo2024iron} appears to show that metal-silicate equilibration causes a measurable amount of iron isotopic fractionation, both at high and low pressures.
This has led to the inference that metal-silicate equilibration may be important for establishing the iron isotopic composition of planetary mantles including the bulk silicate Earth \citep{roskosz2006experimental,polyakov2009iron,shahar2015sulfur, shahar2016iron,elardo2017iron,elardo2019iron,ni2022planet,luo2024iron}.
Depending on the iron alloy composition, thermodynamic conditions of metal-silicate equilibration, and relative abundance of $\text{Fe}^{3+}$, different magnitudes of iron isotopic fractionation have been measured, as shown in Figure \ref{fig:capDelta57Fe_three_panel}.
There are also seemingly conflicting inferred iron isotopic fractionation factors depending on measurement technique and direct comparisons are difficult because experiments are not done with the exact same compositions and thermodynamic states.

The inclusion of alloying elements in the metal may enhance the iron isotopic fractionation due to metal-silicate equilibration, as shown in the left hand panel of Figure \ref{fig:capDelta57Fe_three_panel}.
Some piston cylinder experiments showed a positive correlation between an increase in substitutional impurities in iron metal and the metal becoming isotopically enriched in heavier iron isotopes following metal-silicate equilibration shown in red, orange, blue, and yellow data points in from \citet{elardo2019iron},  \citet{elardo2017iron}, \citet{poitrasson2009iron},
 and \citet{shahar2015sulfur}, respectively.
The magnitude of iron isotopic fractionation in those experiments is $\Delta^{56}\text{Fe}_{\text{sil-met}}=-0.05 \pm 0.22\permil$ at low pressures.
Therefore, core formation is predicted to produce isotopically light mantles, particularly on smaller planets.
If this is the case, then the BSE would be lighter than chondritic, and the heavy signature of mid-ocean ridge basalts (MORBs) would have to be the result of partial melting and not a consequence of metal-silicate equilibration \citep{elardo2017iron, elardo2019iron}.
NRIXS measurements show no trend in fractionation factor due to the presence of substitutional impurities in iron alloy \citet{liu2017iron}, contradicting the many piston-cylinder experiments that show increased Fe-alloy impurities result in metal that is heavier than silicate following equilibration \citet{shahar2020assessment}.
Making inferences from these datasets is further complicated by low pressure experiments performed in \citet{kubik2022absence} which did not find any effect of Ni impurities on the fractionation factor of iron and, furthermore, found minimal iron isotopic fractionation due to metal-silicate equilibration at all, shown in green in the left panel of Figure \ref{fig:capDelta57Fe_three_panel}.
In contrast, \citet{shahar2016iron} finds the silicate phase should be $\sim0.03\permil$ heavier than a pure iron metal after equilibration, and if there is incorporation of H into the metal the silicate phase (bridgmanite) would be $\sim0.07\permil$ heavier than the $\text{FeH}_x$. 
However, this study was performed under high pressure conditions and is thus not directly comparable to other low pressure experiments \citep[e.g.,][]{elardo2019iron, elardo2017iron, poitrasson2009iron, shahar2015sulfur,kubik2022absence}.

Disagreement in prior studies about the ultimate effect of metal-silicate equilibration on iron isotopes in the Earth's mantle stems in part from differing measurement methodologies and also --- with possible significant consequences --- from the interpretation of those measurements within the context of core formation models.
Even if metal-silicate equilibration does fractionate the iron isotopic system a lot or a little, it is not clear how the processes of core formation conspire to maximize or minimize the effects of that fractionation.
Inadequate attention has been given to how isotopic fractionation evolves on a planetary scale: laboratory measurements are typically placed only in the context of single-stage core formation models that often assert complete equilibration between silicate and metal phases within a body \citep[e.g.,][]{roskosz2006experimental,polyakov2009iron,hin2012iron,shahar2015sulfur,shahar2016iron,liu2017iron,elardo2017iron,elardo2019iron,ni2022planet,kubik2022absence,luo2024iron}.
Thus, prior discussions of iron isotopic evolution during core formation are effectively asserting that planetary mantle reservoirs were established in a single equilibration event, which is incompatible with leading theories of planetary accretion and cannot reproduce the moderately siderophile element budget of the Earth's mantle~\citep[e.g.,][]{Rubie2003,Righter2016}.
Notably, all of the prior experimental studies have interpreted their laboratory data in the context of single-stage core formation models, so the potential compounding effects of multiple stages of core formation are neglected.
Here, using numerical models accounting for realistic planetary accretion and differentiation histories, we explore a range of proposed iron isotopic fractionation factors from laboratory experiments.

Modeling Earth's core formation in a single-stage metal-silicate equilibration event is inaccurate and not physically viable for two primary reasons.
First, a single pressure-temperature equilibration requires the whole core and mantle to equilibrate with one another.
However, single stage equilibration can only reproduce siderophile element abundances in the Earth's mantle if equilibration occurs at mid-mantle pressures and temperatures. 
Equilibration at these conditions, as opposed to those of the core-mantle boundary, is inconsistent with whole-body equilibration. 
(Some elements are not even well matched at a single pressure \citep[e.g., Ga and Mn, see][]{mann2009differentiation}, and this difficulty increases with the consideration of additional siderophile elements.)
Instead, multiple stages of varying pressure-temperature conditions must have occurred to explain the moderately siderophile element composition of the bulk silicate Earth (BSE) \citep[e.g.,][]{wanke1981constitution,kegler2008new,mann2009evidence,rubie2015accretion}.
Second, cratering records and inferences from astrophysical planet formation models support that there was extensive impact melting during the growth of the terrestrial planets \citep[e.g.,][]{Wetherill1980,chambers2001planets,walsh2011low,clement2018mars}.
These high-energy impact events would cause melting, material mixing, and subsequent equilibration of silicate and metal reservoirs from impactor and target planetary embryos \citep[e.g.,][]{Tonks1992,Tonks1993,deguen2011experiments,Deguen2014,deVries2016,Nakajima2021}.
Multiple phases of core-formation distinctly shape elemental abundances in a growing Earth-like planet and are necessary to explain the bulk silicate planetary composition \citep[e.g.,][]{rubie2015accretion,rubie2016highly,Fischer2017,Gu2023,nathan2023constraining}.
To adequately understand the behavior of iron isotopes during core formation, modeling multistage accretion and differentiation is necessary, which is the primary motivation for this work.

Assuming a single-stage core formation model means that a small difference in the inferred iron isotopic fractionation factor has significant consequences for the bulk silicate Earth iron composition.
For instance, NRIXS experiments have reported force constants implying limited \citep[e.g.,][]{liu2017iron} to significant \citep[e.g.,][]{ni2022planet} iron isotope fractionation during Earth's core formation.
Specifically, \citet{ni2022planet} consider equilibration between pyrolitic mantle and many different iron alloy compositions (e.g., $\text{Fe}$, $\text{Fe}_{92}\text{Ni}_{8}$, $\text{FeH}_x$, and others) at either the liquidus or solidus of the pyrolitic mantle at a range of pressures between 40 and 60 GPa.
They demonstrate that equilibration between pyrolite and $\text{FeH}_x$ alloys consistently produces the greatest $\Delta^{57}\text{Fe}_{sil-met}$, resulting in the mantle of the Earth being between +$0.03\permil$ and +$0.06\permil$ heavier than chondritic material due to metal-silicate equilibration.
Their measurements and single-stage core formation models further predict a correlation between planetary size and iron isotopic composition; core formation should cause the mantle of small bodies (equilibrating under low-pressure conditions) to have isotopically light iron, while the mantles of larger bodies such as Earth should possess isotopically heavier iron.
\citet{luo2024iron} use \textit{ab initio} calculations to argue that metal-silicate equilibration should result in isotopically heavy iron in the silicate phase at all pressure conditions, because anharmonicity affects the behavior of the metal phase at low pressures.
Notably, the models of core-mantle differentiation in these studies do not consider potential effects of multiple stages of core formation in their assessment of whether metal-silicate equilibration is a plausible driver of iron isotopic fractionation, leaving out a critical modeling consideration that is separate from possible disagreements or evolving understandings of experimentally derived force constants.

Here, we investigate the effect of iron isotopic fractionation due to metal-silicate equilibration during multiple stages of terrestrial accretion and differentiation (based on the model used in \citet{rubie2015accretion, jennings2021metal, nathan2023constraining}).
This terrestrial planet formation model uses inputs of astrophysical N-body simulations of solar system formation and applies geochemical modeling of mantle-core element partitioning following impact-generated melting in order to viably reproduce the astrophysical and geochemical properties of the Earth.
To calculate iron isotopic fractionation during core formation events, we rely on fractionation factors measured by past studies \citep[e.g.,][]{liu2017iron, ni2022planet,luo2024iron}.
With this model, we assess whether or not Earth's iron isotopic composition was meaningfully affected by the process of core-mantle equilibration and separation.

\section{Methods}
\label{methods}

\subsection{Governing equations of iron isotopic fractionation during metal-silicate equilibration}
\label{methods_equations}
Given a degree of isotopic fractionation ($\Delta^{57}\text{Fe}_\text{sil-met}$) between silicate and metal phases of equilibrated metal and silicate material, a system of equations describes how iron isotopes partition.
This system of equations accounts for mass balance during elemental and isotopic partitioning of iron and its isotopes between metal and silicate in an equilibrating melt.  
We consider three isotopes of iron in this system: $^{54}$Fe, $^{56}$Fe, and $^{57}$Fe.
This excludes the other isotopes of iron (e.g. $^{58}$Fe) that make up the complete accounting of Fe; however, $^{54}$Fe, $^{56}$Fe, and $^{57}$Fe  account for >99.7\% of iron isotopes and are sufficient to model this system.
To determine the molar amount of iron isotopes in metal and silicate phases following equilibration, there are six unknown quantities and six equations governing the behavior of the system.

The six unknown quantities in this system are the molar abundance of each isotope of iron ($^{54}\text{Fe}$, $^{56}\text{Fe}$, and $^{57}\text{Fe}$) in either metal ($^{i}\text{Fe}_\text{met}$) or silicate ($^{i}\text{Fe}_\text{sil}$) phase following metal-silicate equilibration for iron isotope $i$: $^{54}\text{Fe}_\text{met}$, $^{54}\text{Fe}_\text{sil}$, $^{56}\text{Fe}_\text{met}$, $^{56}\text{Fe}_\text{sil}$, $^{57}\text{Fe}_\text{met}$, $^{57}\text{Fe}_\text{sil}$.

Six equations determine the behavior of iron isotopes in the metal-silicate equilibration system: three describe the mass balance of the equilibrating Fe, the fourth describes the chemical partitioning of iron between metal and silicate phases, the fifth describes the difference in the $^{57}\text{Fe}/^{54}\text{Fe}$ isotopic composition of the silicate and metal phases of the equilibrated materials, and the sixth describes the relationship between the $^{56}\text{Fe}/^{54}\text{Fe}$ and $^{57}\text{Fe}/^{54}\text{Fe}$ isotopic composition of the equilibrated materials.

\subsubsection*{Mass Balance }
The three equations describing mass balance of iron isotopes in this system are:
\begin{eqnarray}
^{54}\text{Fe}_\text{tot} =\ ^{54}\text{Fe}_\text{met} +\ ^{54}\text{Fe}_\text{sil} \\ 
^{56}\text{Fe}_\text{tot} =\ ^{56}\text{Fe}_\text{met} +\ ^{56}\text{Fe}_\text{sil}  \\ 
^{57}\text{Fe}_\text{tot} =\ ^{57}\text{Fe}_\text{met} +\ ^{57}\text{Fe}_\text{sil}     
\end{eqnarray}
where $^{i}\text{Fe}_\text{tot}$ is the total moles of iron isotope $i$ in the initial equilibrating fluid, and $^{i}\text{Fe}_{met}$ and $^{i}\text{Fe}_{sil}$ are the moles of iron isotope $i$ in the metal and silicate phases, respectively, after equilibration has occurred.

\subsubsection*{Elemental Partitioning }
The molar partition coefficient of iron $D_{\text{Fe}}$ is defined as: 
$$
D_{\text{Fe}} = \frac{X_{\text{Fe}_\text{met}}}{X_{\text{FeO}_\text{sil}}} =
    \frac{\frac{\text{Fe}_\text{met}}{T_\text{met}}}{\frac{\text{FeO}_\text{sil}}{T_\text{sil}}}
$$
where $X_{\text{Fe}_\text{met}}$ and $X_{\text{FeO}_\text{sil}}$ are the molar concentrations of iron in the equilibrating metal and silicate melts.
$\text{Fe}_\text{met} = \sum\limits_{i}\ ^{i}\text{Fe}_\text{met}$ and $\text{FeO}_\text{sil} = \sum\limits_{i}\ ^{i}\text{Fe}_\text{sil}$ are the total moles of iron in either metallic or silicate phase post-equilibration, where $^{i}\text{Fe}_\text{met}$  and $^{i}\text{Fe}_\text{sil}$  are the molar amount of iron isotope $i$ in the equilibrated metal and silicate, respectively.
$T_{\text{metal}}$ and $T_{\text{silicate}}$ are the combined total moles of equilibrating metal and silicate, and these are determined by the details of the equilibrating system under consideration.
$D_{\text{Fe}}$ can be set directly as assessed from high pressure metal-silicate equilibration experiments \citep[e.g.,][]{frost2010partitioning}.

Considering these definitions and the unknown quantities of isotopes in metallic and silicate phase, the fourth equation which governs elemental partitioning between metal and silicate phase is:
\begin{eqnarray}
D_{\text{Fe}} = \frac{\left(^{54}\text{Fe}_\text{met} +\ ^{56}\text{Fe}_\text{met} +\ ^{57}\text{Fe}_\text{met}\right)/ T_\text{metal}}{\left(^{54}\text{Fe}_\text{sil} +\ ^{56}\text{Fe}_\text{sil} +\ ^{57}\text{Fe}_\text{sil}\right)/ T_\text{silicate}}  
\end{eqnarray}

\subsubsection*{Isotopic Fractionation Between Metal and Silicate}
The fifth equation describes the difference in 
$^{57}$Fe isotope composition between the equilibrated silicate and
metallic phases, normalized to the IRMM-014 isotopic standard:
\begin{align}
\Delta^{57}\text{Fe}_\text{sil-met} &= \frac{10^3}{R_{57}} \left(\frac{^{57}\text{Fe}_\text{sil}}{^{54}\text{Fe}_\text{sil}} - \frac{^{57}\text{Fe}_\text{met}}{^{54}\text{Fe}_\text{met}}\right)    
\end{align}

where $R_{57}$ is the $^{57}$Fe/$^{54}$Fe ratio of the
IRMM-014 standard. 
This quantity is the difference in IRMM-014-normalized isotope compositions,
$\delta^{57}\mathrm{Fe}_\mathrm{sil}-\delta^{57}\mathrm{Fe}_\mathrm{met}$, and is used here as the model measure of metal-silicate isotopic fractionation.

The value of $\Delta^{57}\text{Fe}_\text{sil-met}$ is defined by
\begin{equation*}
\Delta^{57}\text{Fe}_\text{sil-met} = \delta^{57}\text{Fe}_\text{sil} - \delta^{57}\text{Fe}_\text{met} 
\end{equation*}
where $\delta^{57}\text{Fe}_\text{phase} $ is the standard delta notation for a given phase.

We considered several different models for the isotopic fractionation factor $\Delta^{57}\text{Fe}_\text{sil-met}$, which are summarized in Table \ref{tab:fractionation_factors}. 
We examine constant fractionation factors as well as pressure- and temperature-independent fractionation factors from two recent NRIXS studies \citet{liu2017iron} and \citet{ni2022planet}, as well as from  \citet{luo2024iron}, which uses an ab-initio approach. 
\citet{liu2017iron} estimates the silicate component (mantle analog) as either bridgmanite or basalt, while
\citet{ni2022planet} estimates the silicate composition with a range of pyrolitic compositions, which is likely more representative of the bulk mantle composition \citep{mcdonough1995composition}.
\citet{luo2024iron} also considers a pyrolitic composition for the mantle analog material.

\begin{table*}[!t]
\centering

\caption{Summary of Fe isotope fractionation factors used in this study.}
\label{tab:fractionation_factors}
\footnotesize
\begin{tabular}{p{2.7cm} p{2.9cm} p{3.2cm} p{3.0cm} p{3.6cm} p{2.0cm}}
\hline
\textbf{Model} & \textbf{Reference} & \textbf{\shortstack{Silicate\\(mantle proxy)}} & \textbf{\shortstack{Metal\\(core proxy)}} & \textbf{Dependence} & \textbf{\shortstack{Relevant\\Figure(s)}}\\
\hline

Constant $\Delta^{57}$Fe$_{\mathrm{sil-met}}$ 
& This study 
& -- 
& -- 
& Fixed values: $-0.1$, $0.01$, $0.05$, $0.10$, $0.2$, $0.3$, $0.4\permil$ 
& Figures \ref{fig:Fe_fractionation_constant_capDelta}, \ref{fig:Liu_versions_vs_constant_fractionation}, \ref{fig:mantle_d57_vs_delta_amount}, \ref{fig:Luo_comparison} \\

Basalt / bridgmanite NRIXS 
& \citet{liu2017iron} 
& Basalt; bridgmanite 
& Fe metal / alloy 
& $P$--$T$ dependent $\Delta^{57}$Fe$_{\mathrm{sil-met}}$ 
& Figure \ref{fig:Liu_versions_vs_constant_fractionation} \\

Pyrolite NRIXS 
& \citet{ni2022planet} 
& Pyrolite with variable Fe$^{3+}$/Fe$_{\mathrm{tot}}$ 
& Fe metal 
& $P$--$T$ dependent $\Delta^{57}$Fe$_{\mathrm{sil-met}}$ 
& Figure \ref{fig:3isotope_nishahar} \\

\textit{Ab initio} pyrolite--Fe melt 
& \citet{luo2024iron} 
& Pyrolite melt 
& Fe melt 
& $P$--$T$ dependent $\Delta^{57}$Fe$_{\mathrm{sil-met}}$ 
& Figure \ref{fig:luo2024} \\

\hline
\end{tabular}
\end{table*}

\subsubsection*{Relationship Between $\delta^{56}\text{Fe}$ and $\delta^{57}\text{Fe}$}
The sixth (and final) equation in this system describes the ratio of isotopic compositions $\delta^{56}\text{Fe}$ and $\delta^{57}\text{Fe}$ in the final equilibrated material:
\begin{eqnarray}
\theta^{56/57} = \frac{\frac{^{56}\text{Fe}_\text{sil}}{^{54}\text{Fe}_\text{sil}} \cdot \frac{1}{R_{56}} - 1}{\frac{^{57}\text{Fe}_\text{sil}}{^{54}\text{Fe}_\text{sil}}\cdot \frac{1}{R_{57}} - 1}  
\end{eqnarray}
where $R_{56}$ and $R_{57}$ are, respectively, the $^{56}$Fe/$^{54}$Fe and $^{57}$Fe/$^{54}$Fe isotopic ratios of the IRMM-014 standard. 
$\theta^{56/57}$ is equal to 0.678, which corresponds to mass-dependent fractionation between $^{56}$Fe and $^{57}$Fe \citep{anbar2000nonbiological,bottrell2003criss}.

\subsubsection*{System of iron isotope equations}
This system of six equations describing iron isotope fractionation is solved for the six unknowns ($^{54}\text{Fe}_\text{met}$, $^{54}\text{Fe}_\text{sil}$, $^{56}\text{Fe}_\text{met}$, $^{56}\text{Fe}_\text{sil}$, $^{57}\text{Fe}_\text{met}$, $^{57}\text{Fe}_\text{sil}$) each time a metal-silicate equilibration event occurs during core formation.
In order to solve these equations, we must know the molar composition of the metallic and silicate liquids undergoing equilibration between the target and projectile bodies, and then from this information and knowledge of the thermodynamic state of the system infer the molar partition coefficient $D_\text{Fe}$ and isotopic fractionation factor $\Delta^{57}$Fe$_{sil-met}$ from high-pressure equilibration experiments.

The outcome of single metal-silicate equilibration events can vary widely due to measured differences of the iron isotope fractionation factor from a significant effect \citep[e.g.,][]{polyakov2009iron,shahar2016iron,elardo2017iron,ni2022planet} to an insignificant effect \citep[e.g.,][]{hin2012iron,liu2017iron}. 
However, the fractionation factor is not the only important factor since the system of equations above depends also upon the elemental partitioning of Fe.
Thus, the behavior of this system of equations is nonlinear, and the post-metal-silicate equilibration event iron isotopic composition is not a simple extrapolation from the laboratory-inferred iron isotope fractionation factor.
Given a fractionation factor $\Delta^{57}$Fe$_{sil-met} = 0.1$ the resulting isotopic composition of equilibrating metal and silicate depends also on the chemical partitioning of Fe, following the lever rule; at lower partition coefficients of Fe, the ratio of iron in silicate compared to metal is closer to unity, while if Fe is more siderophile, more of the fractionation is accommodated by the silicate composition becoming isotopically heavier.

For equilibrium Fe isotope fractionation, the total budget of Fe in metallic and silicate reservoirs is important, not just the partition coefficient of Fe.
This is built into our model, inherited from \citet{rubie2011heterogeneous}; the model does not assume an oxygen fugacity for each equilibration event since the oxygen atoms are directly accounted for in the mass balance of the model. 
This is to say that the FeO content of the Earth evolving throughout accretion is not asserted but instead is a result from self-consistent accounting of mass balance within the growing Earth. 
Each initial body is assigned a starting redox state (i.e., core mass fraction), but after that initial composition, each body's mantle and core are not in full chemical equilibrium.
Instead, the interacting projectile core and part of the target mantle determine the composition of the equilibrating silicate and metallic fluids via mass balance and available oxygen in the accreting bodies.
These fluids are then mixed into the mantle and core reservoirs, respectively.

\subsection{Incorporation into a multistage model of planetary accretion and differentiation}
\label{methods_multistage}
We incorporate the iron isotopic fractionation model described above into a multistage model of planetary accretion and differentiation.
The accretion histories of the terrestrial planets are taken from astrophysical N-body models of solar system formation, and the composition of each body within the solar system formation is tracked using a geochemical model of planetary differentiation.
Accretion events, i.e., planetary impacts, generate melt and mix core material into the target world \citep[e.g.,][]{Tonks1992,Tonks1993}.
After each impact, metal-silicate equilibration occurs between the mantles and cores of the target and impacting planetary bodies in a plume of entrained material surrounding the descending impactor core \citep{Rubie2003}.
Elements partition between the equilibrating mantle- and core-forming fluids from the impactor and target bodies as determined by a mass-balanced calculation that relies on laboratory-measured partition coefficients of material between metal and silicate reservoirs \citep{rubie2011heterogeneous}.
This model is built upon the multistage planetary accretion and differentiation algorithm described originally in \citet{rubie2015accretion}, which successfully reproduces the mantle chemistry of Earth for Earth analogs from an initial protoplanetary disk of chondritic bodies with a radial redox gradient.
This specific multistage model has a rich history \citep[][]{rubie2015accretion,rubie2016highly,Jacobson2017,jennings2021metal,Blanchard2022} but is very similar to other combined accretion-differentiation models in the literature \citep[e.g.,][]{Fischer2017,Gu2023}.

\begin{figure*}[!t]
    \centering

    \begin{subfigure}{0.38\textwidth}
        \centering
        \includegraphics[width=\linewidth]{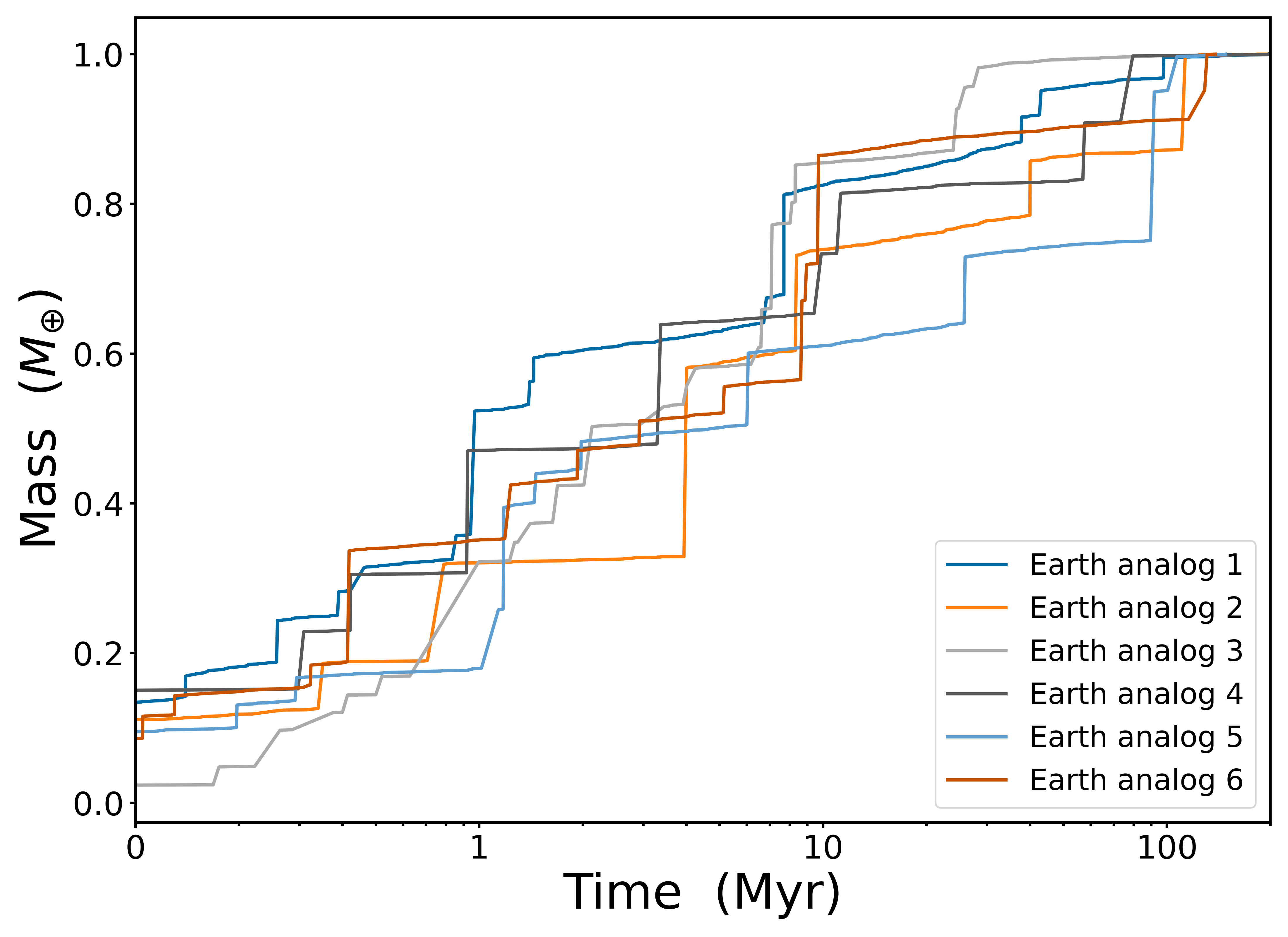}
        \caption{Mass growth histories.}
        \label{fig:EARTH_ANALOGS_MASS_LOGTIME}
    \end{subfigure}
    \begin{subfigure}{0.38\textwidth}
        \centering
        \includegraphics[width=\linewidth]{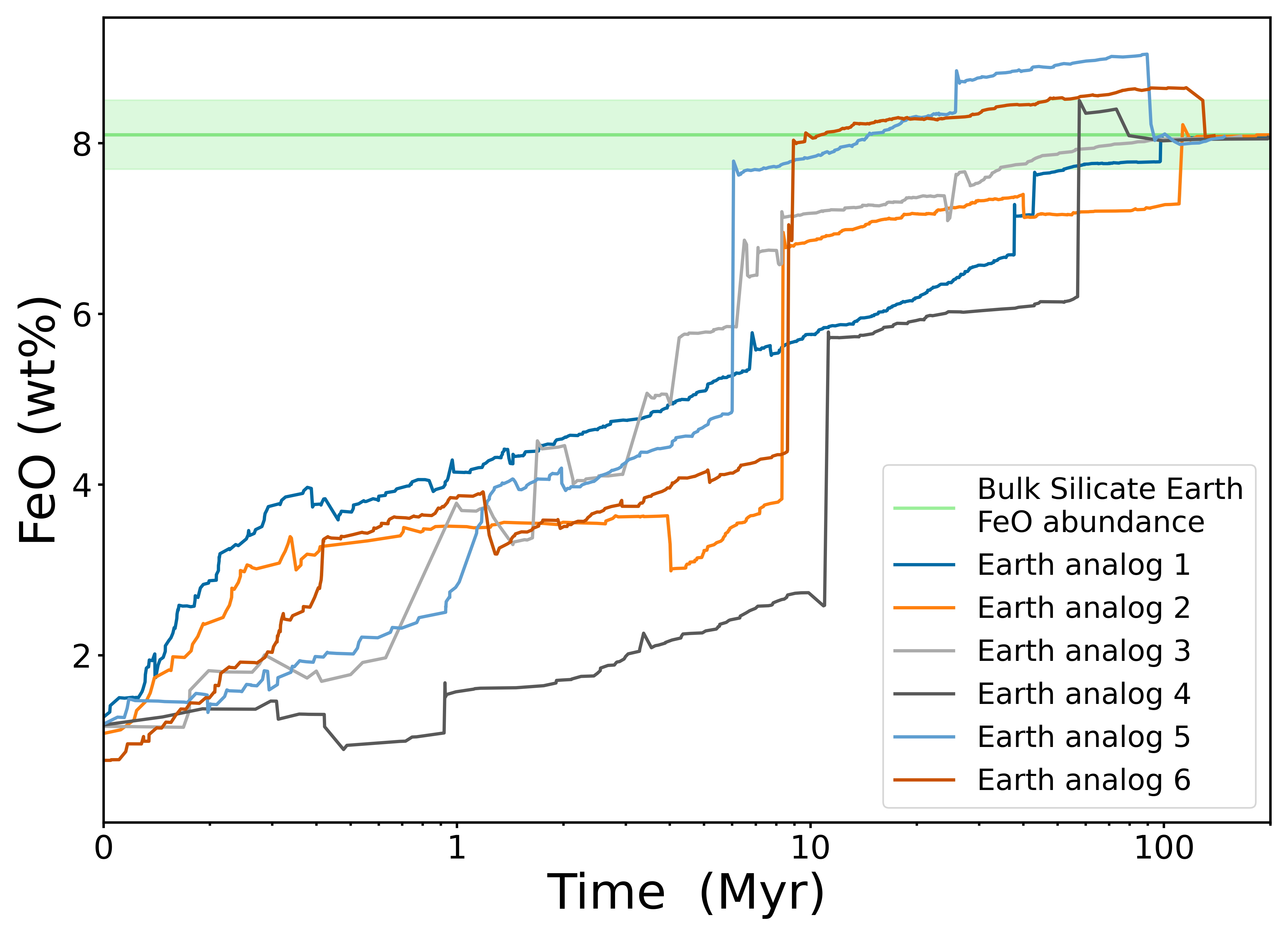}
        \caption{Mantle FeO evolution.}
        \label{fig:FeO_delivery}
    \end{subfigure}
    \caption{
    Growth and mantle FeO evolution of Earth analogs from the six terrestrial planet formation simulations of \protect\citet{jacobson2014lunar}.
    \textbf{(a)} Mass growth histories of Earth analogs as a function of time in millions of years on a logarithmic time scale.
    Each increase in mass represents accretion of a planetesimal, shown as a small increase, or an embryo, shown as a large vertical jump.
    \textbf{(b)} Mantle FeO abundance over time for the same Earth analogs.
    The mantle geochemistry simulations produce mantle FeO compositions that match the bulk silicate Earth, indicated by the green horizontal line and associated uncertainty band from \protect\citet{palme2014cosmochemical}.
    Colors and labels correspond to the same simulations in both panels.
    }
    
    \label{fig:mass_accumulation_and_feo_delivery}
\end{figure*}

\subsubsection*{Accretion histories from astrophysical N-body simulations of terrestrial planet formation}
For this work, the series of accretion events are taken from astrophysical N-body simulations that generally reproduce the astronomical properties of the entire inner solar system, i.e. planetary masses and orbits as well as the mass and orbital distribution of the asteroid belt.
In particular, we use a subset of six terrestrial planet formation simulations from the suite of numerical models presented in \citet{jacobson2014lunar}.
These simulations were shown to feasibly reproduce both Earth's mantle geochemistry \citet{rubie2015accretion} and Mars' mantle geochemistry \citep{nathan2023constraining}.

These astrophysical N-body simulations consider the Grand Tack scenario \citep{walsh2011low,jacobson2014lunar,Brasser2016}, during which the inner terrestrial protoplanetary disk is shaped by the inward-then-outward migration of Jupiter and Saturn.
In this formation scenario, Jupiter and Saturn are in an orbital resonance and migrate inward and then outward, pushing material in the inner solar system closer together resulting in planetesimal and embryo collisions that grow the inner planets.
The simulations were run using the N-body integrator code Symba \citep{duncan1998multiple} and follow the same prescriptions as described in \citet{walsh2011low}.
They were first presented in \citet{jacobson2014lunar} and many more details can be found there.
The Grand Tack scenario is no longer the only scenario that can reproduce the mass-orbit distribution of the terrestrial planets as well as the composition dichotomy of the asteroid belt, and the Grand Tack scenario has uncertainties of its own \citep{Raymond2014,Raymond2020,Raymond2022}.
Alternatives include proposals that the Mars-forming and asteroid belt regions were cleared by an early giant planet instability \citep[e.g.,][]{clement2018mars,clement2019early,Clement2019} or that these regions were always depleted in mass due to structures inherited from the protoplanetary disk itself \citep[e.g.,][]{hansen2009formation,izidoro2015terrestrial,Drazkowska2016,raymond2017empty} and hybrid scenarios \citep[e.g.,][]{Lykawka2023}.
However, the chosen simulations in this work are still broadly representative of successful terrestrial planet formation \citep{raymond2009building,morbidelli2012building,Raymond2020}.

The accretion history of each Earth analog analyzed is consistent with the growth history of Earth as inferred from geochemical evidence \citep[e.g.,][]{jacobson2015earth,Kleine2025}, and this history is typical for terrestrial planets that experience a giant impact era in the presence of a leftover planetesimal disk \citep{raymond2009building,morbidelli2012building,Raymond2020}.
In each of the 6 terrestrial planet formation simulations that we use here, the Earth analog accretes on a timescale of roughly 50--100 million years.  
The mass accretion for each Earth analog in the simulations used in this study is shown in Figure \ref{fig:EARTH_ANALOGS_MASS_LOGTIME}.
These growth histories are consistent with the terrestrial Hf-W chronometer \citep{Kleine2017} and a small late veneer on Earth \citep{jacobson2014highly}.
While the initial locations and masses of the embryos and planetesimals that populate the protoplanetary disk varies between simulations, the example Earth-like planet growth histories used here have similar properties as those from
many successful accretion scenarios \citep[e.g.,][]{chambers2001making,raymond2009building,jacobson2015earth,Raymond2022}.

Considering core formation during multistage accretion, there are two situations when metal-silicate equilibration and corresponding equilibrium isotopic fraction occurs:
\textbf{1) After formation due to runaway growth.} When a planetesimal or embryo undergoes runaway growth, which establishes its initial mass in the astrophysical N-body model, the protoplanet undergoes planetary differentiation.
The runaway growth process is rapid enough and occurs early enough to fully melt the body due to radioactive heating from short-lived radionuclides (e.g., $^{26}$Al) combined with the heat of accretion and the heat released by differentiation itself \citep[e.g.,][]{Elkins-Tanton2017}.
Planetary embryos complete runaway growth to masses between that of the Moon and Mars \citep[e.g.,][]{chambers2001making,morbidelli2012building,jacobson2015earth} but well short of the mass of Earth itself \citep[e.g.,][]{Morbidelli2025,Yap2025}.
Within the planetesimal population, the planetesimal size distribution is dominated by objects with diameters $\gg 100$ km in size \citep[e.g.,][]{Morbidelli2009,Li2019} similar to the modern day asteroid belt \citep{Bottke2015}, so the representative simulated planetesimals would also be large enough to have undergone melting and differentiation \citep{elkins2012magma}.
\textbf{2) After accretion events.} Following a collision between two protoplanets, materials from the mantles and cores of each body will combine in the largest remnant.
Smaller remnant bodies (i.e., debris) may be generated \citep{Agnor2004,Genda2012,Leinhardt2012}, which may affect the relative proportion of mantle and core material in the largest remnant \citep{Asphaug2006,Cambioni2020,Allibert2023,Allibert2025}---a possible hypothesis for the interior of Mercury \citep{Benz1988,Asphaug2014,Chau2018}.
For this study we neglect the role of imperfect accretion and consider that each collision merges the projectile with the target.

The impact energy from the collision generates a melt pool on the larger target, and the metal core of the smaller projectile equilibrates with a fraction of the silicate mantle of the larger target.
In this model, the entire projectile core is smashed into small droplets which equilibrate with a portion of the target mantle at the base of the melt pond.
The amount of silicate melt with which projectile core equilibrates is calculated from \citet{deguen2011experiments}. 
This parameterization typically results in 1 to 10 percent of a target mantle melting and mixing with the projectile core, depending on the impactor size.
These equilibrating reservoirs combine separately with the non-equilibrating reservoirs: the freshly equilibrated metallic liquids with the pre-existing target core and the freshly equilibrated silicate melt with the target mantle and accreted projectile mantle.

From the astrophysical N-body simulations and geochemical box modeling, each simulation in this study produces an astrophysical Earth analog (mass and orbit) with a mantle composition that matches the abundance of refractory lithophile and moderately siderophile elements estimated to compose the bulk silicate Earth across a suite of elements (Mg, Ca, Al, Na, Fe, Si, O, Ni, Co, V, Cr, Nb, and Ta), similar to \cite{rubie2015accretion} and \citet{nathan2023constraining}.
We use the best-fit parameters describing the initial oxidation state in the protoplanetary disk and pressures of equilibration as found in \citet{nathan2023constraining} to produce a simulated Earth mantle composition that closely resembles the bulk silicate Earth (BSE) composition (BSE defined by \citet{palme2014cosmochemical}).
Because the model uses a mass-balance approach to calculating metal-silicate equilibration, the redox state of the accreting planets evolves naturally based on the materials involved in each equilibration event. 
Thus, the core size and FeO content are not merely applied to the growing Earth, but are instead a result of self-consistent record keeping of the chemical composition in a simulated Earth mantle.
See \citet{nathan2023constraining} for model details and the procedure for finding the best-fit parameters.
Indeed, these models match the modal FeO abundance of the mantle, see Figure \ref{fig:FeO_delivery}, which shows the abundance of iron oxide in the mantle of the six simulated Earth-like planets as a function of time.

\section{Results: iron isotopic fractionation during multistage accretion}
\label{results_multistage}

\begin{figure*}[!t]
    \centering


    \begin{subfigure}[t]{0.38\textwidth}
        \centering
        \includegraphics[width=\linewidth]{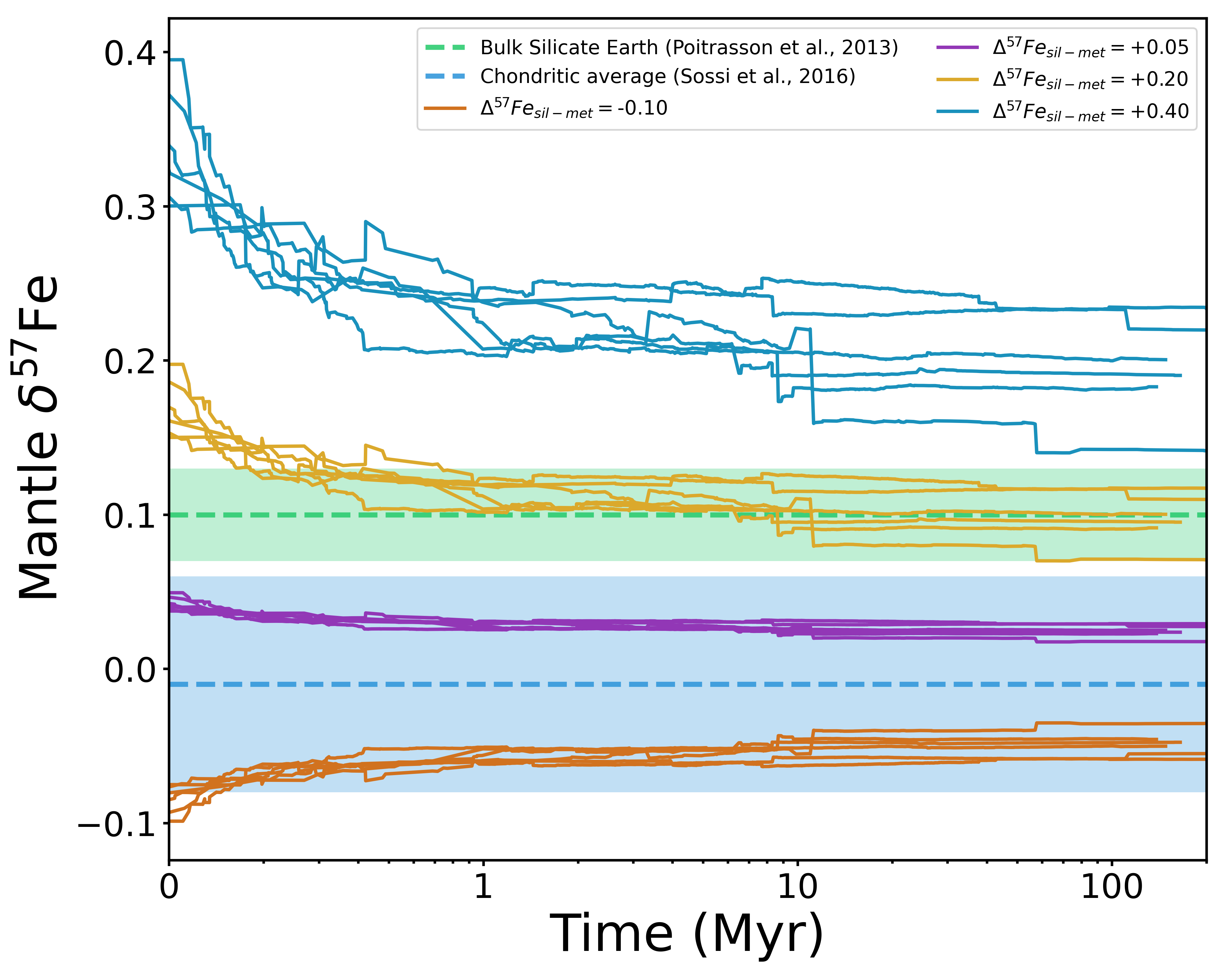}
        \caption{Constant fractionation factors.}
        \label{fig:Fe_fractionation_constant_capDelta}
    \end{subfigure}
    \begin{subfigure}[t]{0.38\textwidth}
        \centering
        \includegraphics[width=\linewidth]{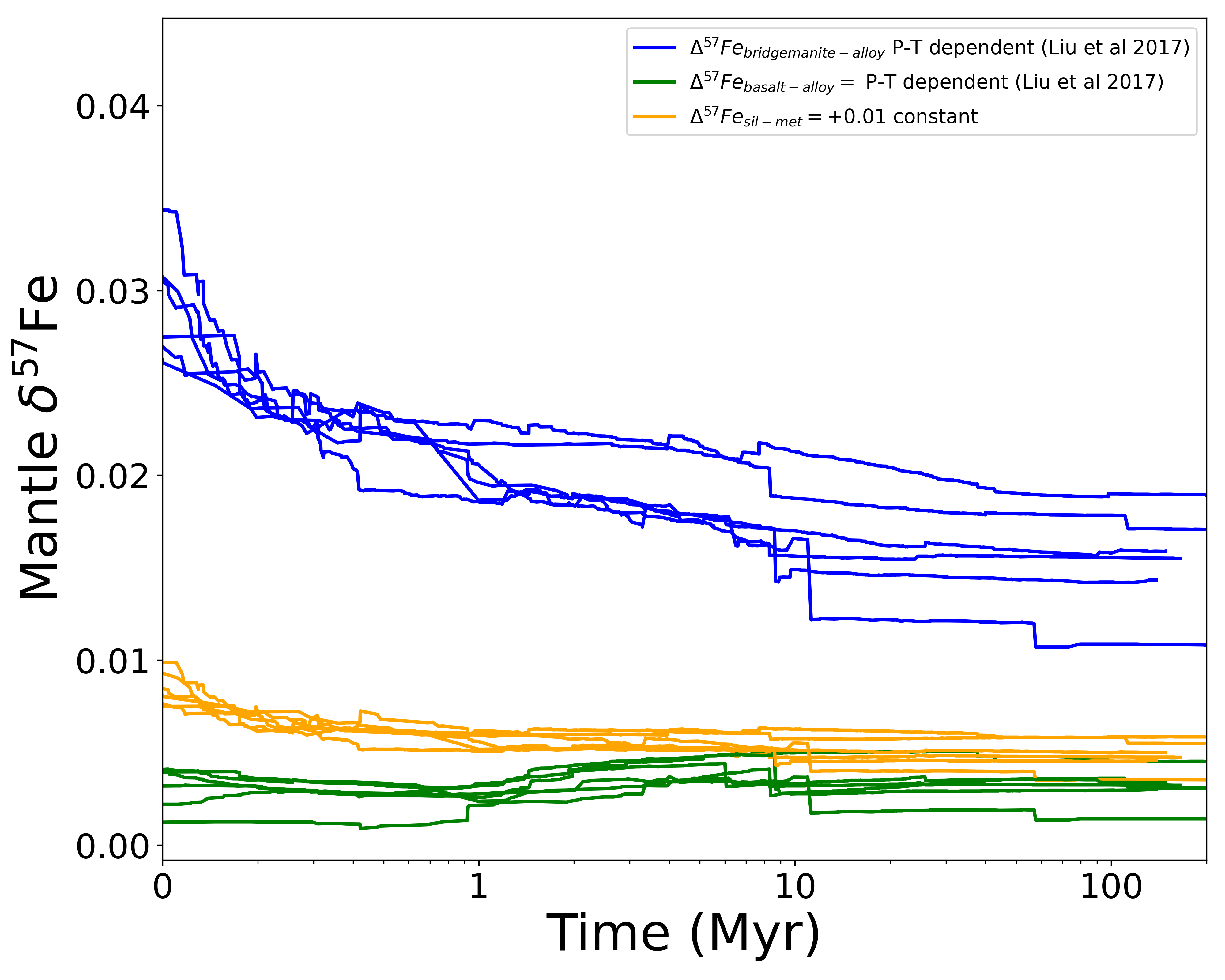}
        \caption{\protect\citet{liu2017iron} fractionation factors.}
        \label{fig:Liu_versions_vs_constant_fractionation}
    \end{subfigure}

    \vspace{1em}


    \begin{subfigure}[t]{0.38\textwidth}
        \centering
        \includegraphics[width=\linewidth]{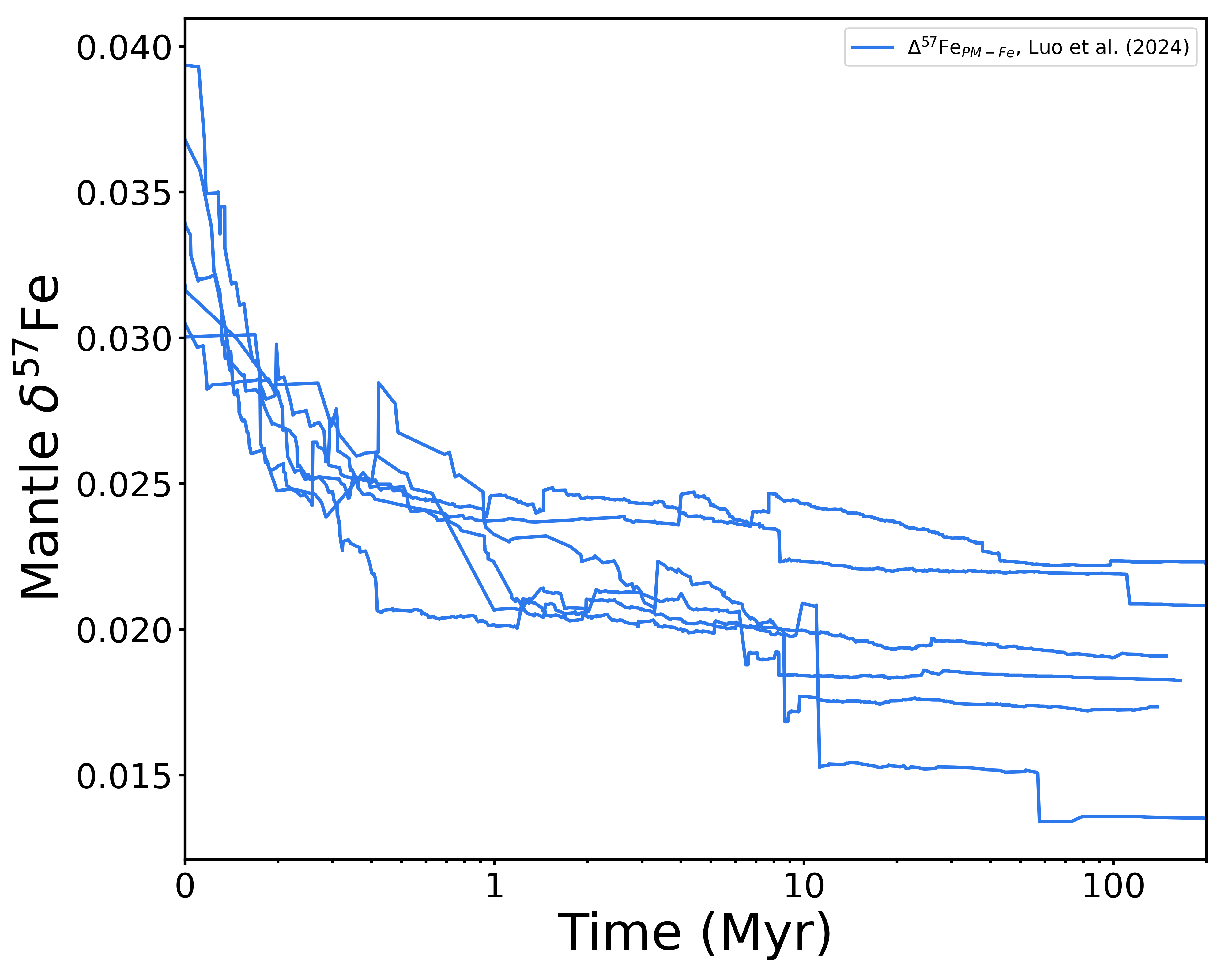}
        \caption{\protect\citet{luo2024iron} fractionation factors.}
        \label{fig:luo2024}
    \end{subfigure}

    \captionsetup{
        justification=justified,
        singlelinecheck=false
    }

    \caption{
    Modeled mantle $\delta^{57}\text{Fe}$ evolution for different iron isotope fractionation prescriptions.
    \textbf{(a)} Constant, temperature- and pressure-independent isotopic fractionation factors are applied during metal-silicate equilibration.
    Each colored group of lines represents the set of six terrestrial planet formation simulations, with colors corresponding to different magnitudes of constant $\Delta^{57}\text{Fe}_\text{sil-met}$.
    Shaded regions indicate solar system reference values and their reported uncertainties.
    \textbf{(b)} Pressure- and temperature-dependent fractionation factors from \protect\citet{liu2017iron} are compared with constant fractionation.
    The bridgmanite analog produces heavier final mantle compositions than the basalt-alloy case, while constant $\Delta^{57}\text{Fe}_\text{sil-met}=+0.01\permil$ gives intermediate values.
    \textbf{(c)} Fractionation factors based on \protect\citet{luo2024iron} produce final mantle enrichments of approximately $+0.01\permil$ to $+0.025\permil$, depending on stochastic accretion history.
    }
    \label{fig:three_fractionation_models}

\end{figure*}

\begin{figure*}[!t]
    \centering

    \includegraphics[width=0.58\textwidth]{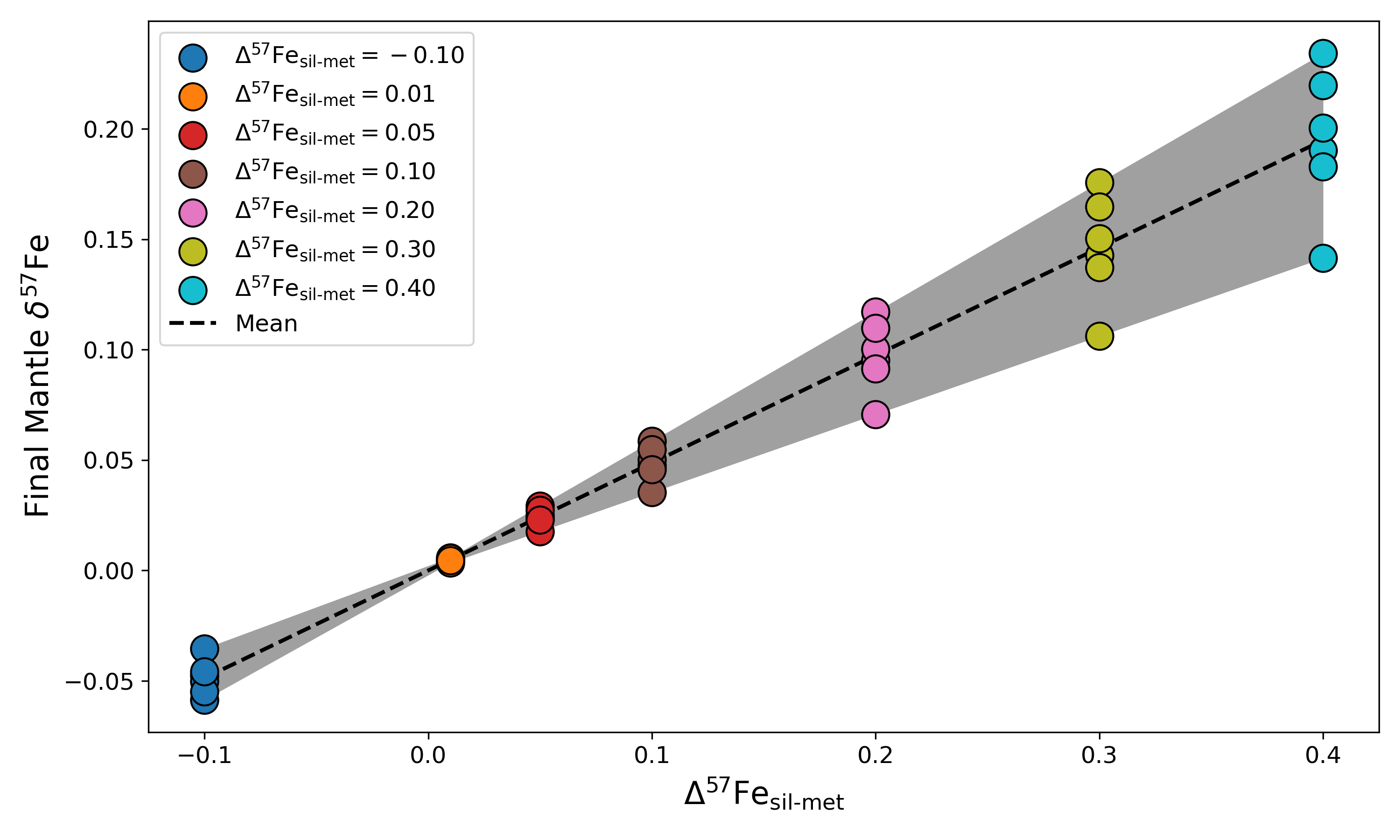}

    \caption{
    Mantle $\delta^{57}\text{Fe}$ of simulated Earth-like planets as a function of constant fractionation factor $\Delta^{57}\text{Fe}_{\text{sil-met}}$ is shown.
    Each marker represents one of the six Grand Tack simulations' Earth-like planet.
    Markers are colored by the value of constant fractionation factor applied throughout simulated accretion.
    The dashed line and shaded region represent the mean and range of calculated mantle iron isotopic composition for a given set of simulations and fractionation factor.
    }
    \label{fig:mantle_d57_vs_delta_amount}
\end{figure*}

\subsection{Pressure and temperature independent iron isotope fractionation factor}
\label{constant_fractionation}
In order to isolate the role of multistage core formation, we first apply a constant fractionation factor $\Delta^{57}\text{Fe}_\text{sil-met}$ for iron isotopic fractionation within the simulated planets.
We experiment with a range of constant values from $\Delta^{57}\text{Fe}_\text{sil-met}=-0.10\permil$ to $\Delta^{57}\text{Fe}_\text{sil-met}=+0.40\permil$.
Note that we assumed large fractionation factors that are comparable with the  $\delta^{57}$Fe composition of several notable solar system bodies (BSE, CM Chondrite, Moon, Mars), but these assumed isotopic fractionation factors are generally much larger than those measured in the laboratory.
In each case, when the planetary embryo undergoes planetary differentiation after runaway growth at the start of the simulation, the mantle of the embryo establishes the $\delta^{57}$Fe composition consistent with the assumed fractionation factor and initial bulk composition of the embryo.
Following each impact and melting event, the iron in the equilibrating material isotopically fractionates at a given constant value between the metal and silicate phases, according to the mass balance constraints described in Section \ref{methods}.
Subsequent accretion changes the average iron isotopic composition of the growing embryo mantle over the course of accretion, as shown in Figure \ref{fig:Fe_fractionation_constant_capDelta}.

Across all the simulations and fractionation factors tested, the isotopic composition trends back towards the assumed chondritic building block composition of $0.0\permil$ over multiple accretion and subsequent core formation events (see evolving trends in Figure \ref{fig:Fe_fractionation_constant_capDelta}).
This rebalancing effect is a result of mass balance between the equilibrating material accommodating the transfer of heavy isotopes between metal and silicate reservoirs.
This material is then mixed with the rest of the mantle which did not melt during impact, thus did not experience a fractionation factor and stayed at its previous composition.
Repeated equilibration events amplify these two effects, rebalancing and mixing, and result in a mantle composition that trends towards chondritic over the course of accretion.
This means that the magnitude of the iron isotopic fractionation of the Earth's mantle is always smaller than the magnitude of the metal-silicate fractionation factor it undergoes during equilibration, as shown in Figure \ref{fig:mantle_d57_vs_delta_amount}.

The trend towards chondritic isotopic composition despite fractionation events occurs because of the addition of the impactor core --- which tends to be fractionated in an opposite direction to the fractionation of the target mantle --- and also because not all of the target mantle melts and chemically re-equilibrates.
The isotopically equilibrated fluids are strongly influenced by the isotopic compositions of the impactor core, which fully equilibrates. 
If the projectile core did not equilibrate and instead sequestered to the target core immediately, this effect would be lessened, and in this way there is a dependence on model construction.
Additionally all of the mantle of the target body (and none of its core) equilibrates with the projectile after impact and this non-fractionating component of the mantle re-mixes following the impact.
This combined effect reduces the net fractionation that occurs during equilibration.

Rebalancing and small fractionation factors greatly reduce the likelihood that, even in a multistage core formation scenario, $\delta^{57}$Fe is meaningfully affected by metal-silicate equilibration during core formation, and also demonstrates the value of considering an astrophysically motivated model of core formation.
Repeated episodes of core formation continue to drive the mantle isotopic composition towards a value slightly different from a chondritic starting point, but significantly less in magnitude than the fractionation factor assumed to produce an elevated mantle value by prior studies.

In these scenarios, a range of $\Delta^{57}\text{Fe}_\text{sil-met}$ could theoretically result in the BSE iron isotopic composition.
For instance, a constant $\Delta^{57}\text{Fe}_\text{sil-met} = +0.1\permil$ best produces an Earth mantle iron isotopic composition closest to that of the Earth.
However, as we show in the subsequent sections, laboratory measured fractionation factors are pressure dependent and, as a result, have significantly smaller effects on the fractionation of the Earth's mantle.

\subsection{Pressure and temperature dependent isotopic fractionation of iron using either a bridgmanite or a basalt model}
\label{dependent_fractionation}
Pressure and temperature dependent fractionation factors cause the mantle $\delta^{57}$Fe composition to vary in less straightforward ways than constant fractionation over the course of accretion, but show similar trends towards the chondritic baseline.
Naturally, laboratory measurements require a physical model for the silicate melt equilibrating with the iron metal.
\citet{liu2017iron} used bridgmanite and basalt as model systems.
While bridgmanite is a much better model system for Earth's mantle, we examine both here for completeness.
The pressure and temperature dependent fractionation factors $\Delta^{57}\text{Fe}_\text{sil-met} $ for bridgmanite and basalt mantle composition models were obtained from force constants extracted via NRIXS \citep{liu2017iron}.
Both the bridgmanite model and the basalt model use force constants for a metal component with composition  $\text{Fe}_{86.8} \text{Ni}_{8.6} \text{Si}_{4.6}$, as also measured in \citet{liu2017iron}.

Both the bridgmanite and basalt models of mantle composition are dependent on temperature and pressure; however, the final iron isotopic fractionation for both cases is minimal.
The results of these two models are shown in Figure \ref{fig:Liu_versions_vs_constant_fractionation}, compared with a constant fractionation factor of $\Delta^{57}\text{Fe}_\text{sil-met}=+0.01\permil$ for context.
The fractionation factors inferred from NRIXS measurements for bridgmanite-metal alloy produce a final mantle $\delta^{57}$Fe between $\sim$+0.01 $\permil$ and +0.02 $\permil$, depending on the particular N-body accretion simulation in use.
Whereas, the iron isotope fractionation factors of the basalt-metal alloy system produce a final mantle $\delta^{57}$Fe  between $\sim$0.0 $\permil$ and +0.005 $\permil$, dependent again on the particular N-body accretion simulation.
Applying a constant $\Delta^{57}\text{Fe}_\text{sil-met}$ of $+0.01\permil$ (i.e., pressure and temperature independent) produces values intermediate between the bridgmanite and basalt models.
Thus, we find that in a multistage core accretion scenario, very little iron isotopic fractionation occurs using the measured fractionation factors from \citet{liu2017iron}, regardless of the silicate model system used, bridgmanite or basalt.
While part of this result can be attributed to the small measured fractionation factor, the rebalancing effect first noted in the constant isotopic fractionation case essentially removes the isotopic fractionation signal in this case.

\subsection{Effects of $\text{Fe}^{3+}/\sum{\text{Fe}}$ variation on Fe fractionation }
\label{pyrolite_fractionation}
Next, we considered the effects of varying $\text{Fe}^{3+}/\sum\text{Fe}$ ratios in a pyrolitic silicate on  the evolution of iron isotopic composition in the mantle.
\citet{ni2022planet} used NRIXS to obtain isotopic fractionation factors for iron isotopes partitioning between metal and a pyrolitic silicate with a range of $\text{Fe}^{3+}/\sum\text{Fe}$ ratios.
\citet{ni2022planet} considers equilibration with pure iron with force constants measured in \citet{murphy2013experimental}.
The isotopic fractionation factor $\Delta^{57}\text{Fe}_\text{sil-met}$ between a silicate with pyrolite composition and pure iron varies with pressure, temperature, and the $\text{Fe}^{3+}/\sum\text{Fe}$ ratio of the pyrolite.
For example, at 3500 K and $\text{Fe}^{3+}/\sum\text{Fe}=0.0$, the fractionation factor $\Delta^{57}\text{Fe}_\text{sil-met}$ ranges from $\sim-0.01$ at low pressure (<10 GPa) to $\sim+0.04$ at high pressure (60 GPa).
For a ratio $\text{Fe}^{3+}/\sum\text{Fe}=1.0$, where all iron in pyrolite is $\text{Fe}^{3+}$, the behavior of the fractionation factor is reversed; the fractionation factor $\Delta^{57}\text{Fe}_\text{sil-met}$ ranges from $\sim+0.04$ at low pressure (0 GPa) to $\sim-0.03$ at high pressure (60 GPa).
We note that at low pressures $\sim$1 bar, iron metal cannot be in equilibrium with silicate melt if $\text{Fe}^{3+}/\sum\text{Fe}=1$, making the low pressure range of the fractionation for assemblages with $\text{Fe}^{3+}/\sum\text{Fe}>0$ a modeling exercise and not physically viable. 
Instead, experiments suggest that the $\text{Fe}^{3+}/\sum\text{Fe}$ should increase from $\sim 0$ at low pressure and increase to             $\sim 0.2$ at higher pressures \citep{Armstrong2019magmaocean, zhang2024ferric}. 

\begin{figure*}[!t]
    \centering
    \includegraphics[width=0.65\textwidth]{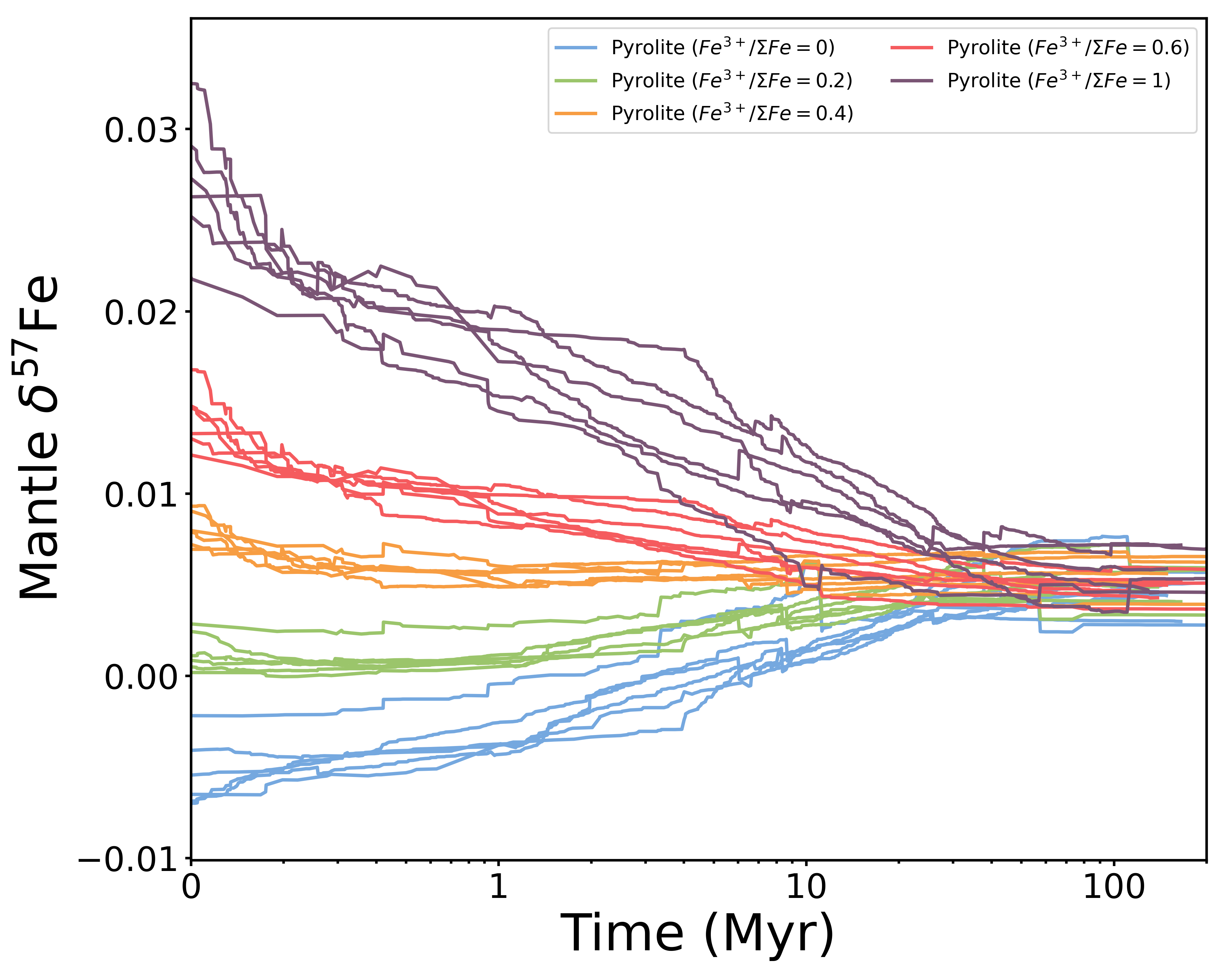} 
    \caption{$\delta^{57}\text{Fe}$ mantle composition of simulated Earth analogs with application of pressure-dependent and pyrolite composition-dependent iron fractionation factors measured in \protect\citet{ni2022planet}. Groups of lines represent the six simulations used in this study, each color represents a different fractionation factor corresponding to a different oxidation state of pyrolite representing the mantle composition. Mantle $\delta^{57}\text{Fe}$ composition is shown on the y-axis changing with respect to time in millions of years shown in logarithmic scale on the x-axis. The groups of simulations, regardless of specific pyrolitic composition, all converge to a near-chondritic value ($\sim0.0\permil$) due to the effects of repeated impacts and remelting during multistage accretion and differentiation. }
    \label{fig:3isotope_nishahar}
\end{figure*}

In the multistage formation scenario simulations, the fractionation factors for the range of $\text{Fe}^{3+}$ pyrolitic compositions from \citet{ni2022planet} ultimately do not affect the iron fractionation during metal-silicate equilibration.
Multiple impacts and instances of re-equilibration cause all the separate pyrolitic compositions to trend towards the same final iron isotopic composition, similar to the effects observed in the first tests shown in the previous section.
The results of this are shown in Figure \ref{fig:3isotope_nishahar} which shows that the final mantle $\delta^{57}$Fe composition is between $+0.005\permil$ and $+0.01\permil$, regardless of the pyrolite composition.
This is because the range of force constants for these compositions range from negative to positive over the pressure and temperature ranges of Earth formation.
Initial fractionation is closely associated with pyrolitic composition but the different compositions trend towards the same value over the course of a longer ($\sim$100 Myr) Earth accretion.
The effect of different pyrolitic compositions is negated via longer accretion histories that evolve over a range of PT conditions.
The final value is still dependent on the accretion history but not predictably so.
The final composition is also not measurably different than a scenario without fractionation.

In \citet{ni2022planet}, the measured pressure and temperature dependence of iron isotopic fractionation is presented as evidence for the mantles of small planetary bodies having isotopically light iron composition following core formation and those of larger bodies like Earth having isotopically heavy iron composition.
\citet{ni2022planet} predicts a super-chondritic iron composition of the Earth's mantle due to isotopic fractionation during metal-silicate equilibration between $+0.03\permil$ and $+0.06\permil$.
However, this is a much greater magnitude of isotopic fractionation than that predicted by a multistage core formation scenario.
When we consider the effects of rebalancing and mixing with non-equilibrating reservoirs, there is a convergence that is due to the range of pressure-temperature regimes achieved during planetary formation.

\subsection{Fractionation factors from Luo et al. (2024)}
\label{luo_section}
Lastly, \textit{ab initio} estimates from \citet{luo2024iron} show a greater magnitude of iron isotopic fractionation at low pressures than predicted by NRIXS in \citet{ni2022planet}.
We tested these fractionation factors as well, to see if a greater magnitude of fractionation at low pressures would have an effect.
To obtain full ranges of fractionation factors, we performed a linear fit to measured fractionation factors ($\Delta^{57}\text{Fe}_\text{sil-met}$) between pyrolite melt and liquid iron as reported in Figure 2 of \citet{luo2024iron}; the result is a nearly constant fractionation factor of around $\sim+0.04\permil$, which we apply to the model.
When we apply \citet{luo2024iron} fractionation factors in a simulated accretion and differentiation scenario, the Earth mantle Fe isotopic composition evolves similarly to when \citet{liu2017iron} fractionation factors are applied.
This is shown in Figure \ref{fig:luo2024}.

\begin{figure*}[!t]
    \centering
    \includegraphics[width=0.6\textwidth]{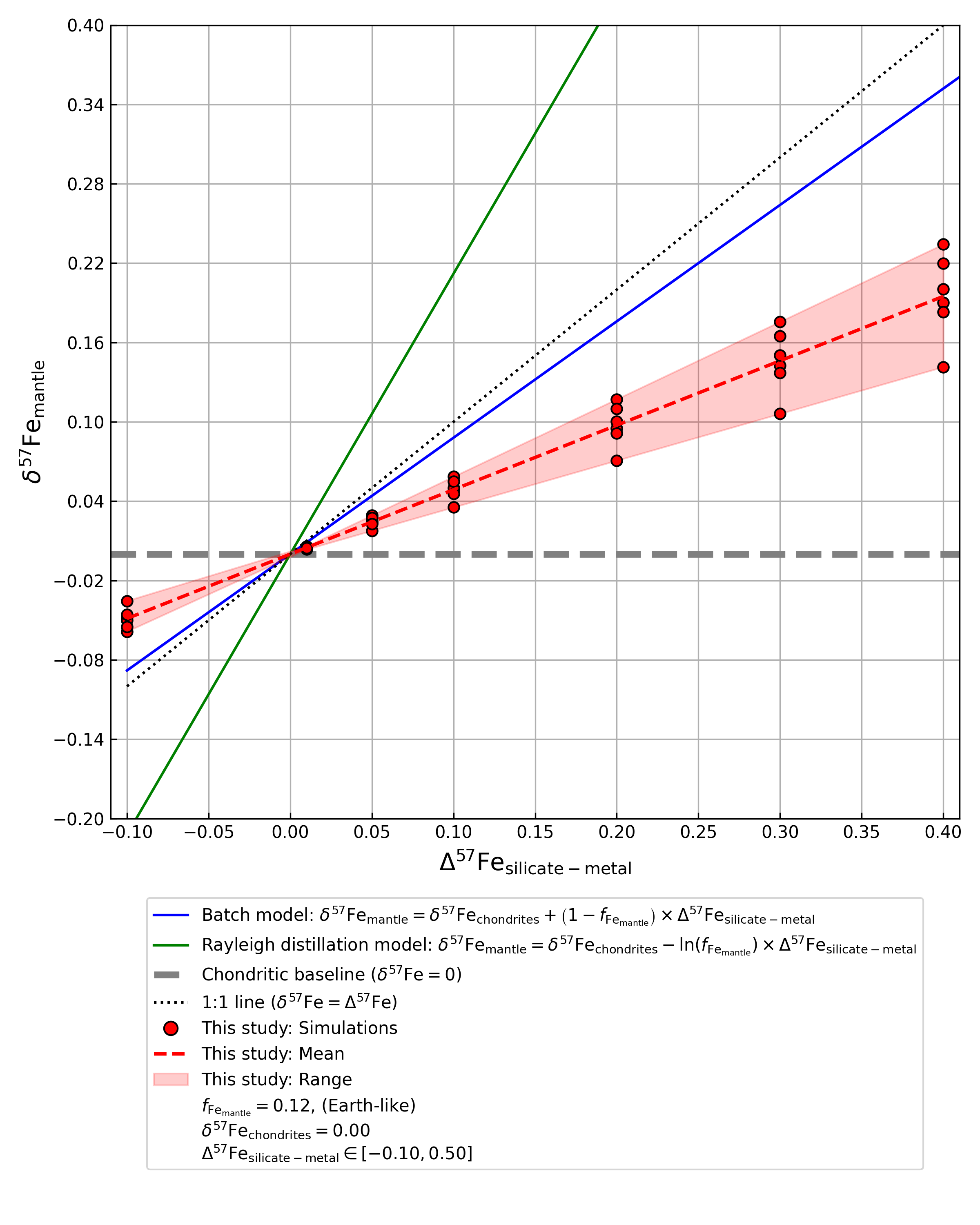}
    \caption{Comparison of the relationship between fractionation factor ($\Delta^{57}\text{Fe}_{\text{sil-met}}$) and mantle iron isotopic composition ($\delta^{57}\text{Fe}_{\text{mantle}}$) of the Batch and Rayleigh models in \citet{luo2024iron} with the calculated mantle of Earth-like planets in this study. Blue and green lines represent the Batch and Rayleigh distillation model relationships, respectively. The dotted black line represents a 1-to-1 relationship between fractionation factor and mantle isotopic composition. Red markers indicate individual simulations at specific $\Delta^{57}\text{Fe}_{sil-met}$ fractionation factors. The dashed red line and shaded region represent the mean and range of calculated mantle iron isotopic composition for the set of simulations and fractionation factors considered here. The gray dashed line represents a chondritic baseline of $\delta^{57}\text{Fe} = 0.0\permil$. 
    \label{fig:Luo_comparison}}
\end{figure*}

Our model predicts a smaller magnitude of iron isotopic fractionation in the mantle of Earth $\delta^{57}\text{Fe}_\text{mantle}$ for a given fractionation factor $\Delta^{57}\text{Fe}_\text{sil-met}$ and Earth-like fraction of total Earth Fe in the mantle $f_{\text{Fe}_\text{mantle}}\approx 0.12$ than is predicted by either of the fractionation models in \citet{luo2024iron}, as shown in Figure \ref{fig:Luo_comparison}.
When comparing the results of the model undergoing constant fractionation with both fractionation models (batch and rayleigh) presented in \citet{luo2024iron}, we again find a reduced fractionation effect.
This reduced fractionation factor is shown empirically to be $\sim$~0.5, as represented by the slope of the mean of plotted values in Figure \ref{fig:Luo_comparison}.
Though there is stochastic variation due to specific accretion histories, this is the average reduction in mantle isotopic fractionation for an Earth-like body.
Other factors, such as degree of mantle equilibration during core formation may have an effect but are secondary to the rebalancing process described.

\section{Discussion}
Here, we have modeled iron isotopic evolution in the Earth during multistage accretion to determine whether core formation has had a measurable effect on the iron isotopic composition of the mantle of planetary bodies. 
We explored iron isotopic fractionation during core formation in the following five scenarios: 1) constant iron isotopic fractionation factors (Section \ref{constant_fractionation}), 2) pressure- and temperature-dependent fractionation factors considering basaltic and 3) bridgmanitic approximations of mantle composition (Section \ref{dependent_fractionation}), 4) fractionation factors based on a pyrolitic mantle composition (Section \ref{pyrolite_fractionation}), and 5) fractionation factors based on \textit{ab initio} calculations of equilibration between pyrolite melt and iron melt from \citet{luo2024iron} (Section \ref{luo_section}).
In each, similar trends appeared: multiple stages of core formation resulted in minimal iron isotopic fractionation and the final mantle isotopic fractionation was smaller in magnitude than the applied fractionation factor due to dilution from non-equilibrating materials and mass balance between metal and silicate reservoirs.

We do not explore the full range of fractionation factors presented in the literature, instead focusing on a set of studies' estimates of fractionation during metal-silicate equilibration that might most clearly determine whether or not core formation significantly affected the mantle Fe isotopic composition of the Earth. 
For example, we have chosen to not include the fractionation conditions in \citet{shahar2015sulfur}; this study finds large fractionation effects when there is a large (up to 25\%) amount of substitutional impurities alloyed with iron.
However, when net fractionation is greatest in \citet{shahar2015sulfur}, it is the metal phase that is heavier than the silicate phase which would make the Earth mantle lighter than chondritic, not heavier.
We were primarily interested in testing scenarios where Earth’s mantle could become heavier because observations indicate Earth’s mantle is heavier than chondritic.
Similarly, though they are displayed in Figure 2a, we also did not model fractionation factors from \citet{kubik2022absence} because that study itself concluded that inclusion of Ni in alloy with Fe had no effect on isotopic fractionation of Fe.

Our results show that there is minimal to no effect of metal-silicate equilibration during core formation on the final iron isotopic composition of a planetary body due to its core formation history.
Even if such fractionation occurred, it at most caused a shift of less than $0.01\permil$ in the isotopic composition of the Earth's mantle over the course of formation.
This magnitude of fractionation is so small that we cannot attribute it coherently to a process like core formation that will have a stochastic range of final isotopic compositions associated with the particular number and intensity of impacts experienced by a growing body.
Inductively coupled plasma mass spectrometry (ICP-MS) measurements of iron isotopic composition routinely demonstrate precision within $0.03\permil$ for $\delta^{56}$Fe \citep{he2015high, dauphas2009routine} and $0.05\permil$ for $\delta^{57}$Fe \citep{craddock2011iron}.
The variation we predict here is even smaller than the precision capabilities of those measurements.

We note that the specific accretion histories of the Earth analogs presented in the simulations here are not particularly important, nor are they the driving factor of the results.
Rather than specific events (such as precise heliocentric origin, mass, and particular impacts experienced by each building block of Earth), the most important factor in the compositional evolution of the Earth mantle is an extended history of accretion and repeated core formation events (and associated changing regime of core formation pressures, temperatures, and fugacity).
Regarding effects of fugacity, we note that there is an increase in the FeO content over time in the Earth, as shown in Figure \ref{fig:FeO_delivery}. 
This trend means that, by the lever rule which controls much of the rebalancing effect we describe, fractionation is largest in magnitude in the silicate phase at earlier times in the Earth's accretion and smaller later in accretion.
Dependence on the stochastic nature of planetary accretion histories remains, which always presents a challenge untangling deterministic outcomes from those of chance.
With the set of solar system formation simulations we examine here, we see core formation is not a driving force in producing iron isotopic fractionation in the Earth's mantle. 
While we believe these results are generalizable to all classical planetesimal accretion scenarios for the Earth, it is plausible that the accretion pathway we modeled is not representative of the Earth's formation and this is the reason we see minimal effects in these experiments.

Metal-silicate equilibration during terrestrial core formation does not appear to be a viable mechanism to have caused iron isotopic fractionation between Earth's mantle and core.
If the bulk Earth iron isotopic composition is heavier than chondrites, then other potential mechanisms must be the source of solar system iron isotopic heterogeneity.
One possible source of isotopic heterogeneity is inherited heterogeneity from the protoplanetary disk.
The Earth was likely built from isotopically distinct materials from other solar system bodies \citep[e.g.,][]{burkhardt2021, dauphas2017isotopic}.
However, despite a spread in isotopic compositions, when compared on a three isotope plot, iron isotopic compositions of all solar system materials (terrestrial and meteoritic) rest on a single fractionation line, implying a degree of homogenization of the initial reservoir in the presolar nebula \citep{zhu2001isotopic}.
Indeed, recent work shows nearly indistinguishable Earth and Moon iron isotopic composition according to nucleosynthetic anomalies in $\mu^{54}\text{Fe}$ and $\mu^{58}\text{Fe}$ \citep{hopp2025inner} implying accretion from a similar reservoir.
Further, we note that nucleosynthetic iron isotopic variation is of a much smaller magnitude than variation in $\delta^{57}\text{Fe}$ \citep{hopp2026iron}.

If the bulk Earth inherited a chondritic iron isotope composition and core formation does not isotopically fractionate the silicate and metallic compositions, then other planetary-scale processes may have governed the iron isotopic composition of silicate materials on Earth.
Two additional hypotheses to explain why silicate Earth materials are offset from a chondritic composition include:
(1) Planetary impacts may have caused vaporization of iron in planetary impactor cores \citep{kraus2015impact} and the resultant vapor loss may have driven equilibrium isotopic fractionation \citep[e.g.,][]{sossi2016ironsystematics,hin2017,young2019,sossi2021shahar}.
(2) Disproportionation within the mantle between $\text{Fe}^{2+}$ and $\text{Fe}^{3+}$ could result in a signature of fractionation that explains the heavy iron isotopic composition of terrestrial basalts \citep{williams2012oxidation}.
However, given the very small amount of $\text{Fe}^{3+}$ as a fraction of total iron in the upper mantle, the degree of isotopic fractionation driven by disproportionation is likely very minimal \citep{craddock2013abyssal}.

We anticipate the rebalancing effect we have documented here to be similar in other isotopic systems, particularly if those elements are delivered to the Earth in multiple episodes of impacts and melting throughout accretion.
It is unclear whether core formation can ever have a significant (or traceable) effect on the isotopic composition of the mantle of a large planetary body.

\section{Conclusions}
Using numerical simulations of terrestrial planet formation, we demonstrated that multiple stages of planetary accretion and core-mantle differentiation did not meaningfully affect the iron isotopic composition of the growing Earth.
Our key findings are the following:
\begin{enumerate}
    \item Laboratory-measured iron isotopic fractionation factors due to metal-silicate equilibration are too small to result in meaningful iron isotopic fractionation of the Earth's mantle, especially in scenarios where the Earth experienced multiple episodes of re-melting and equilibration during its formation.
    \item Our results are independent of the many varied literature values of the fractionation factor $\Delta^{57}\text{Fe}_\text{sil-met}$, measured with a range of experimental techniques.
    \item This is due to reduction in the effective magnitude of metal-silicate equilibration during an episode of impact-induced core formation: mass balance between the metal and silicate reservoirs of the equilibrating fluid results in a mantle fractionation $|\delta^{57}\text{Fe}_\text{mantle}|$ that is less than the magnitude of the fractionation factor $|\Delta^{57}\text{Fe}_\text{sil-met}|$. This rebalancing effect is compounded by multiple episodes of core formation that involve incomplete equilibration of the mantle and re-mixing with non-fractionated material from the impactor.

\end{enumerate}

Given the effects of multiple impact-driven core formation events, the formation of a terrestrial planet core tends to result in a terrestrial mantle with an isotopic composition close to that of its building blocks.

\section*{Acknowledgments}
Research reported in this publication was supported in part by funding provided by the Michigan Space Grant Consortium (MSGC), under award number 80NSSC20M0124 from the National Aeronautics and Space Administration (NASA), and in part from the Future Investigators in NASA Earth and Space Science and Technology (FINESST), under award number 80NSSC23K1376 from NASA. 
This work was also supported in part through computational resources and services provided by the Institute for Cyber-Enabled Research at Michigan State University.

\section*{Data Availability}
All data, code, and inputs relevant to this study are publicly available through Zenodo at   
\href{https://doi.org/10.5281/zenodo.21729566}{10.5281/zenodo.21729566}.  
The Zenodo archive mirrors the GitHub repository  
\url{https://github.com/gijnathan/Fe_isotope_GCA}.

\bibliographystyle{cas-model2-names}
\bibliography{biblio}

\end{document}